\documentclass{article}
\usepackage{microtype}
\usepackage{graphicx}

\usepackage{booktabs} 

\usepackage{hyperref}

\usepackage{caption}
\usepackage{subcaption}
\usepackage[accepted]{icml2021}

\usepackage{amsmath,amsfonts,bm}

\def\eqref#1{equation~\ref{#1}}

\def\1{\bm{1}}

\DeclareMathAlphabet{\mathsfit}{\encodingdefault}{\sfdefault}{m}{sl}
\SetMathAlphabet{\mathsfit}{bold}{\encodingdefault}{\sfdefault}{bx}{n}

\def\gD{{\mathcal{D}}}

\def\gI{{\mathcal{I}}}

\def\gP{{\mathcal{P}}}

\def\gZ{{\mathcal{Z}}}

\def\sI{{\mathbb{I}}}

\def\sR{{\mathbb{R}}}

\newcommand{\E}{\mathbb{E}}

\usepackage{hyperref}
\usepackage{url}
\usepackage{algorithm}
\usepackage{algorithmic}
\usepackage{amssymb}
\usepackage{amsmath}
\usepackage{lipsum}

\newtheorem{assumption}{Assumption}

\newtheorem{lemma}{Lemma}

\usepackage{thmtools}
\usepackage{thm-restate}
\usepackage{hyperref}
\usepackage{cleveref}

\renewcommand{\algorithmicrequire}{\textbf{Input:}} 
\renewcommand{\algorithmicensure}{\textbf{Output:}}

\newcommand{\ci}{\!\perp\!}

\icmltitlerunning{T-SCI: A Two-Stage Conformal Inference Algorithm with Guaranteed Coverage for Cox-MLP}

\begin{document}

\twocolumn[
\icmltitle{T-SCI: A Two-Stage Conformal Inference \\ Algorithm with Guaranteed Coverage for Cox-MLP}




    
    

\icmlsetsymbol{equal}{*}

\begin{icmlauthorlist}
\icmlauthor{Jiaye Teng}{equal,to}
\icmlauthor{Zeren Tan}{equal,to}
\icmlauthor{Yang Yuan}{to}
\end{icmlauthorlist}

\icmlaffiliation{to}{IIIS, Tsinghua University, China}

\icmlcorrespondingauthor{Yang Yuan}{yuanyang@tsinghua.edu.cn}

\icmlkeywords{Machine Learning, ICML}

\vskip 0.3in
]

\newcommand\blfootnote[1]{%
\begingroup
\renewcommand\thefootnote{}\footnote{#1}%
\addtocounter{footnote}{-1}%
\endgroup
}



\begin{abstract}
It is challenging to deal with censored data, where we only have access to the incomplete information of survival time instead of its exact value.
Fortunately, under linear predictor assumption, people can obtain guaranteed coverage for the confidence band of survival time using methods like Cox Regression.
However, when relaxing the linear assumption with neural networks (e.g., Cox-MLP \citep{katzman2018deepsurv,kvamme2019time}), we lose the guaranteed coverage.
To recover the guaranteed coverage without linear assumption, we propose two algorithms based on conformal inference under strong ignorability assumption.
In the first algorithm \emph{WCCI}, we revisit weighted conformal inference and introduce a new non-conformity score based on partial likelihood.
We then propose a two-stage algorithm \emph{T-SCI}, where we run WCCI in the first stage and apply quantile conformal inference to calibrate the results in the second stage.
Theoretical analysis shows that T-SCI returns guaranteed coverage under milder assumptions than WCCI.
We conduct extensive experiments on synthetic data and real data using different methods, which validate our analysis.
\end{abstract}
\section{Introduction}
\begin{figure}[t]
    \centering
    \includegraphics[width=8cm]{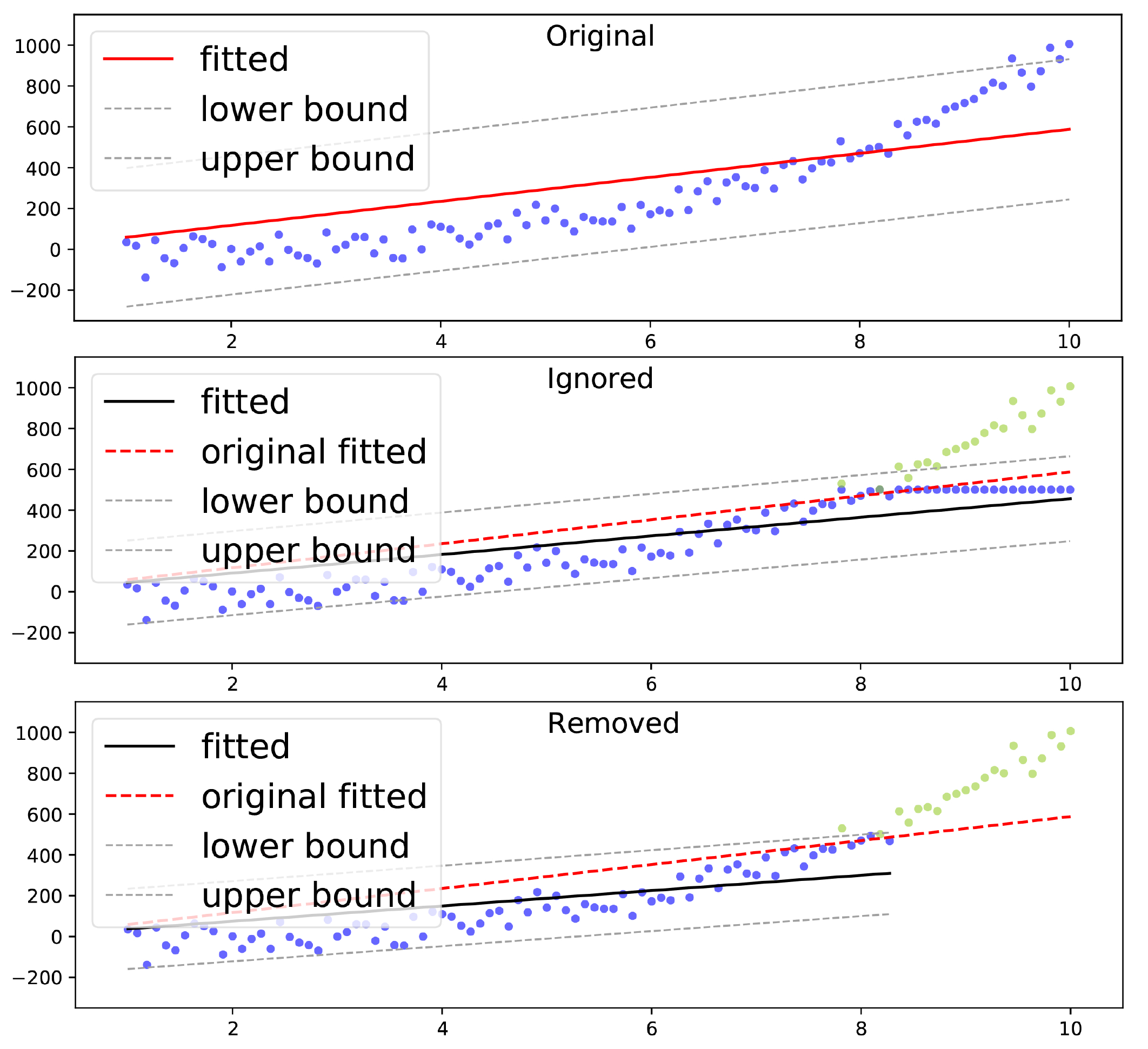}
    \caption{\textbf{Bias in censoring.} Ignoring or deleting censored data (light green points) causes bias compared to the ground truth linear approximation (red line), where the linear approximation under censoring (black line) does not overlap the ground truth (red line).}
\label{fig: biasefficiency}
\end{figure}

In survival analysis, censoring indicates that the value of interest (survival time) is only partially known (e.g., the information can be $t>5$ instead of $t=7$).
\blfootnote{$^*$Equal contribution \ $^1$IIIS, Tsinghua University, China. Email: \{tjy20, tanzr20\}@mails.tsinghua.edu.cn. Correspondence to: Yang Yuan $<$ yuanyang@tsinghua.edu.cn $>$.}
It is common and inevitable in numerous fields, including medical care \citep{robins2000correcting,klein2006survival}, astronomy \citep{feldmann2019leo}, finance \citep{bellotti2009credit}, etc.
It is an annoying issue since ignoring or deleting censored data causes bias and inefficiency \citep{nakagawa2008missing}, as illustrated in Figure~\ref{fig: biasefficiency}.

When dealing with censored data, we usually focus on the \emph{confidence band} of the survival time since confidence bands give a more conservative estimation than point estimation.
Under linear assumptions on covariate effect (See Assumption~\ref{assump: separableeffect}), one can derive the survival time distribution using Cox regression by asymptotic normality of linear coefficient.
It further leads to guaranteed coverage, meaning that survival time provably falls into the confidence band with high probability (larger than the given confidence level).

However, the linear assumption broadly harms its performances and restricts its applications.
After all, the reality is not always entirely linear. 
To relax the linear assumption, \citet{katzman2018deepsurv,kvamme2019time} applies neural networks into Cox regression (Cox-MLP), yielding the best performance in terms of some metrics such as Brier score and binomial log-likelihood.
Unfortunately, the confidence band in Cox-MLP has no guaranteed coverage since one cannot expect neural network \emph{converges} to the expected function, not to mention the asymptotic normality.

In general, we can use \emph{conformal inference} to recover the confidence band with guaranteed coverage by splitting the dataset into training and calibration set \citep{vovk2005algorithmic,nouretdinov2011machine,lei2020conformal}.
One of its advantages is that conformal inference does not harm the model performance since it is post-hoc.
Therefore, we propose to apply conformal inference into Cox-MLP.

When applying conformal inference into Cox-MLP, there are several problems to be solved.
Firstly, Cox regression does not return survival time explicitly, which requires a modification of the non-conformity score.
Secondly, censoring causes \emph{covariate shift} under strong ignorability, meaning that the covariate distribution differs in censored and uncensored data.
Therefore, we cannot apply conformal inference directly.
Thirdly, we need to consider the estimation error and provide theoretical guarantees for the coverage.

In this paper, we propose a new non-conformity score under strong ignorability assumption (which is a standard assumption in weighted conformal inference. We refer to more details in Section~\ref{Sec: preliminary}) based on the partial likelihood of Cox regression.
This non-conformity score does not need an explicit estimation of the survival time.
We then apply weighted conformal censoring inference (WCCI), a weighted conformal inference based on this non-conformity score inspired by \citet{tibshirani2019conformal} to deal with the covariate shift problem.
Furthermore, inspired by \citet{romano2019conformalized}, we provide a two-stage conformal inference (T-SCI) which returns ``nearly perfect'' coverage, meaning that the coverage has not only guaranteed lower bound but also upper bound.
Inspired by \cite{lei2020conformal}, we provide theoretical guarantees for both WCCI and T-SCI algorithms.

We summarize our contributions as follows:
\begin{itemize}
    \item We provide coverage for Cox-MLP in WCCI based on weighted conformal inference frameworks by introducing a new non-conformity score. 
    \item We further propose a T-SCI algorithm based on the quantile conformal inference framework. We show that T-SCI returns nearly perfect guaranteed coverage, namely, theoretical guarantees for coverage's lower and upper bound.
    \item We conduct extensive experiments on both synthetic data and real-world data, showing that the T-SCI-based algorithm outperforms other approaches in terms of empirical coverage and interval length. 
\end{itemize}
\vspace{-2pt}

\section{Related work}

\textbf{Censored data analysis.} An early analysis of censored data can be dating back to the famous  Kaplan–Meier estimator \citet{kaplan1958nonparametric}.
However, this approach is valid only when all patients have the same survival function.
Therefore, several individual-level analysis is proposed, such as proportional hazard model \citep{breslow1975analysis}, accelerated failure time model \citep{wei1992accelerated} and Tree-based models \citep{zhu2012recursively,li2020censored}.

On the other hand, researchers apply machine learning techniques to deal with censored data \citep{wang2019machine}.
For example, random survival forests \citep{ishwaran2008random} train random forests using the log-rank test as the splitting criterion.
Moreover, DeepHit \citep{lee2018deephit} apply neural networks to estimate the probability mass function and introduce a ranking loss.
This paper mainly focuses on the \emph{Cox-based model}, one of the famous proportional hazard model branches.

\textbf{Cox regression} was first proposed in \citet{cox1972regression}, which is a semi-parametric method focusing on estimating the hazard function.
Among all its extensions, \citet{akritas1995theil} first proposed using a one-hidden layer perceptron to replace the linear predictor of the coefficient. 
However, it generally failed mainly due to the low expressivity of the one-hidden layer perceptron \citep{xiang2000comparison,sargent2001comparison}. 
Therefore, \citet{katzman2018deepsurv} proposed to use the multi-layer perceptron instead of the one-layer perceptron (DeepSurv).
Furthermore, \cite{kvamme2019time} generalize the idea to the non-proportional hazard settings.
In this paper, we unify their names as \emph{Cox-MLP} when the context is clear.
However, this line of work lacks a theoretical guarantee.
This paper tries to fill this blank and propose the first guaranteed coverage of the survival time.


\textbf{Conformal inference} was pioneered by Vladimir Vovk and his collaborators [e.g., \citet{vovk2005algorithmic,shafer2008tutorial,nouretdinov2011machine}], focusing on the inference of response variables by splitting a training fold and a calibration fold.
There are several variations of conformal inference.
For example, weighted conformal inference \citep{tibshirani2019conformal} focus on dealing with the covariate shift phenomenon, and quantile conformal inference \citep{romano2019conformalized} returns coverage with not only an upper bound but also the lower bound guaranteed coverage.
Conformal inference, as well as its variations, are widely studied and used in \citet{lei2013distribution,lei2014distribution,lei2018distribution,barber2019limits,sadinle2019least,romano2020classification}.

Recently, \citet{lei2020conformal} apply conformal inference under counterfactual settings and derive a double robust guarantee for their proposed methods.
Our work is partially inspired by \citet{lei2020conformal} but considers a different censoring setting.
Furthermore, we propose a different algorithm and derive a ``nearly perfect'' guaranteed coverage.
A very recent work \citep{candes2021conformalized} focuses on a similar censoring scenario.
Unlike our approach (T-SCI), \citet{candes2021conformalized} relax the strong ignorability assumption but require all information of censoring time to obtain confidence bands.
We emphasize that it is still an open problem on deriving guaranteed confidence bands under general censoring scenarios.

There are some approaches to apply conformal inference under censoring settings.
For example, \citet{bostr2017conformal,bostrom2019predicting} apply conformal inference into random survival forests, and \citet{chen2020deep} derive the confidence band for DeepHit. 
However, one cannot apply these approaches to Cox-based models since Cox-based models do not explicitly return a predicted survival time.


\section{Preliminary}
\label{Sec: preliminary}

\newcommand{\xv}{X}
\newcommand{\yv}{Y}
\newcommand{\cv}{C}
\newcommand{\iv}{\Delta}
\newcommand{\tv}{T}

\newcommand{\xvi}{X_i}
\newcommand{\yvi}{Y_i}
\newcommand{\cvi}{C_i}
\newcommand{\ivi}{\Delta_i}
\newcommand{\tvi}{T_i}

Denote $\xv \sim \gP_X \subset \mathbb{R}^{d}$ as the covariate, $\tv \sim \gP_T \subset \sR$ as the survival time, and $\cv \sim \gP_C \subset \sR$ as the censoring time.
Denote the joint distribution as $\gP_{XTC}$, its marginal distribution as $\gP_{TX}$, and its conditional distribution as $\gP_{T|X}$. 
The survival time cannot be observed when it is larger than the censoring time. Therefore, observed time is the minimum of censoring time and survival time. Let $\yv \in \sR$ be the observed time with uncensoring indicator $\iv \in \sR$, then 
\begin{equation*}
\begin{split}
    \yv = \min \{\tv, \cv\},\ \iv = \sI_{\{ \tv \leq \cv \}}.
\end{split}
\end{equation*}
Denote the dataset as $\gZ = \{ \xvi, \yvi, \ivi \}_{{i \in \gI}}$ where we can only access the covariate, the observed time, and the censoring indicator.
However, the value of interest is the survival time $\tv$.
Therefore, the censored data is incomplete due to the information loss when $\iv = 0$.
We next introduce how censoring happens, namely, the censoring mechanism.

\textbf{Censoring Mechanism.}
In this paper, we consider the censoring regimes with strong ignorability assumption, namely
$\tv \ci \iv \ | \ \xv$.
We will further discuss the strong ignorability assumption in the supplementary materials.

\begin{figure}
    \centering
    \includegraphics[width=0.96\linewidth]{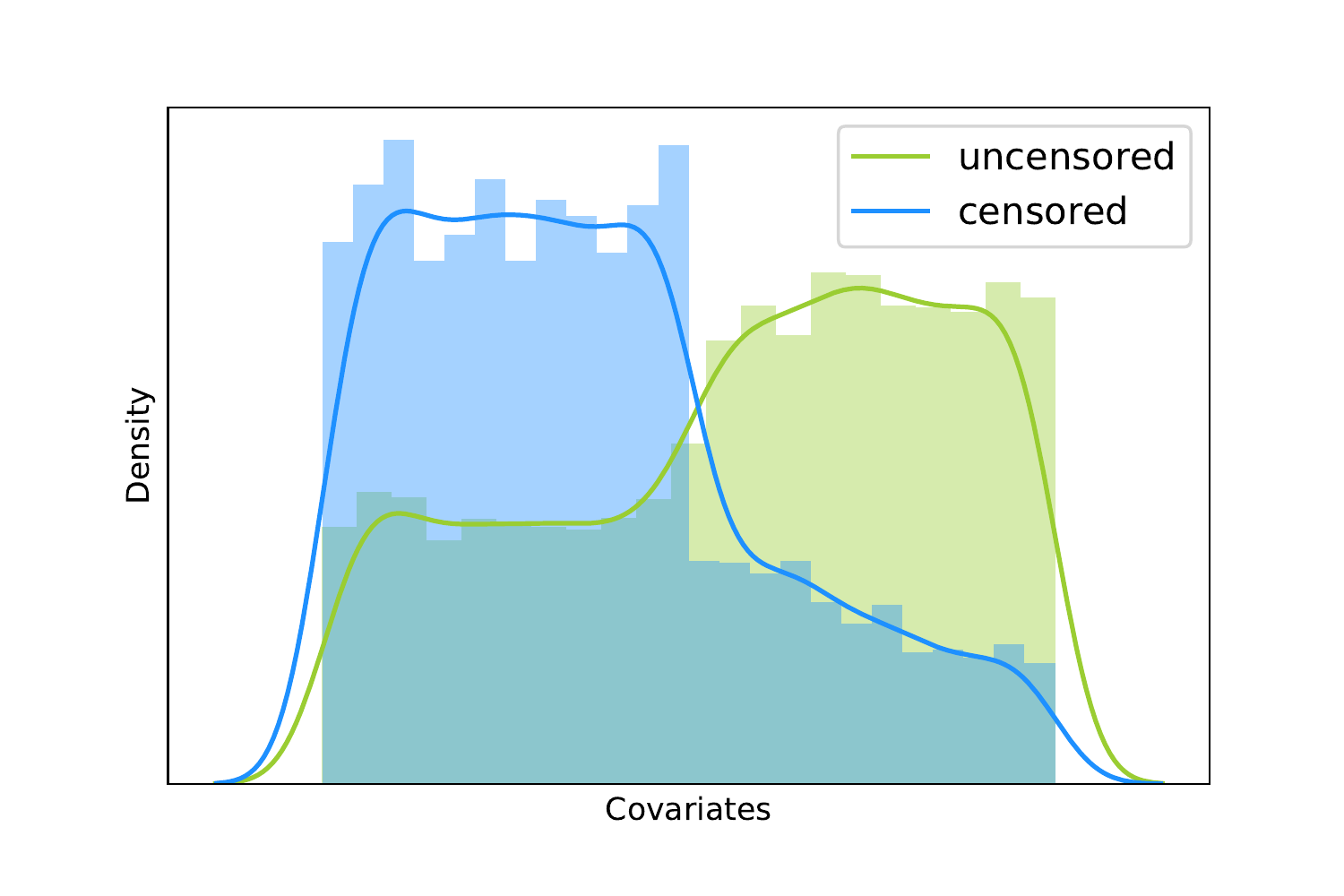}
     \caption{\textbf{Covariate shift illustration.} We show distributions of $x_1$ with (blue) and without (green) censoring under simulation data. Obviously, censored and uncensored data have different distributions, i.e. covariate shift\protect\footnotemark[1].}
    \label{fig: covariate shift}
\end{figure}

However, note that there can be covariate shift under such censoring regimes, namely, the distributions of covariate $\xv$ under censoring and non-censoring are different
$$\left(\xv \ | \ \iv = 1 \right) \overset{d}{\not=}  \left(\xv \ |\ \iv = 0\right).$$
We refer to Figure~\ref{fig: covariate shift} for an illustration. 
Usually, we estimate the distribution of the survival time via the dataset $\gD$ by \emph{Cox regression}. 
We denote $F_\tv(t)$ as the survival time's CDF.

\textbf{Cox Regression.}
In Cox regression, we focus on two important terms 
\emph{survival function} $S_\tv(t)$ and \emph{cumulative hazard function} $\Lambda_\tv(t)$, as defined in Equation~\ref{eqn: survivalfunction}.
We emphasize that they are defined with respect to the survival time $\tv$ instead of the observed time $\yv$.
\begin{equation}
\label{eqn: survivalfunction}
\begin{split}
S_\tv(t) \triangleq 1-F_\tv(t), \ \Lambda_\tv(t) \triangleq -\log S_\tv(t).
\end{split}
\end{equation}
When the context is clear, we omit the subscript $T$ and denote the above function as $F(t)$, $S(t)$, and $\Lambda(t)$, respectively. 
Cox-based models usually require \emph{proportional hazard} assumption,
formally stated in  Assumption~\ref{assump: separableeffect}.

\begin{assumption}[Proportional Hazard] 
\label{assump: separableeffect}
For each individual $i$, we assume
\begin{equation}
\label{Eqn: separable}
\Lambda_i(t; \xvi) = \Lambda_0(t)  \exp \left(g(\xvi)\right)
\end{equation}
where $\Lambda_0(t)$ is the baseline cumulative hazard function, and $g(\xvi)$ is the individual effect named as \emph{predictor}.
\end{assumption}

\textbf{Remark:} There are several non-proportional hazard Cox models which replace $g(x_i)$ with $g(x_i, t)$ (e.g., \citet{kvamme2019time}). Although our proposed algorithm can be directly generalized to non-proportional settings, we only consider proportional hazard models for clarity in this paper.

Specifically, Cox regression solves the case when the predictor $g(\cdot)$ is linear by maximizing \emph{partial log-likelihood} $l(g)$, defined in Equation~\ref{eqn: partial likelihood}.
\begin{equation}
\label{eqn: partial likelihood}
l(g) \triangleq \sum_{j: \Delta_j=1} \log \left(\sum_{k \in R(T_j)} \exp [g(X_k) - g(X_j)]\right),
\end{equation}
where $R (T_j)$ is the set of all individuals at risk at time $T_{j}-$ (the observed time is no less than $T_{j}$), and $g(X_j)=X_j^\top \beta$ is the linear predictor.



For the case when $g(\cdot)$ is not linear, 
\citet{lee2018deephit} and \citet{kvamme2019time} propose Cox-MLP which uses neural networks to replace the linear predictor $g(\xvi)$.
Concretely, they use the negative partial log-likelihood as the training loss, with a penalty on the complexity of $g(\cdot)$.

\footnotetext[1]{The simulation dataset is from \citet{kvamme2019time}}

\textbf{Conformal inference.}
Conformal inference does post-hoc estimation based on the splitting of training and calibration fold.
One trains a model $\mu$ using the training fold and then calculates the \emph{non-conformity score} $V_i$ on the calibration fold.
A commonly used non-conformity score is the absolute error $V_i = |T_i - \mu(X_i)|$ where $(X_i, T_i)$ is the $i_{th}$ calibration sample.
When designing the non-conformity score, one of the most important characteristics is \emph{exchangeability}.

\begin{assumption}[Exchangeability]
\label{Exchangeability}
For $n\geq 1$ random variables $V_1, \dots, V_n$, they satisfy exchangeability if 
$$
(V_1, \dots, V_n) \overset{d}{=} (V_{\pi(1)}, \dots, V_{\pi(n)})
$$
for any permutation $\pi: [n] \to [n]$, where $\overset{d}{=}$ means they have the same distribution, and $[n] = \{1, 2, \dots, n\}$.
\end{assumption}

Exchangeability is weaker than independence since independence implies exchangeability.
Under the exchangeability assumption on the non-conformity score, we reach a theoretical guarantee of the {confidence band}.
In this paper, we mainly focus on two varieties of conformal inference, \emph{weighted conformal inference} (to deal with covariate shift, Lemma~\ref{WCI}) and \emph{quantile conformal inference} (to return a nearly perfectly guarantee, Lemma~\ref{QCI}), respectively.
\begin{lemma}[WCI, \citet{tibshirani2019conformal} Theorem~2]
\label{WCI}
\textbf{\emph{(Informal.)}} Assume that the data $(X_i, T_i), i \in [n+1]$ are weighted exchangeable with weight $w_1, \dots, w_{n+1}$, then the returned confidence band $\widehat{C}_n$ has guaranteed coverage:
$$
\mathbb{P}(T_{n+1} \in \widehat{C}_n(X_{n+1})) \geq 1-\alpha.
$$
\end{lemma}

\begin{lemma}[QCI, \citet{romano2019conformalized} Theorem~1]
\label{QCI}
\textbf{\emph{(Informal.)}} Assume that the data $(X_i, T_i), i \in [n+1]$ are exchangeable, and the non-conformity scores are almost surely distinct, then the returned confidence band  $\widehat{C}_n$ is nearly perfectly calibrated:
$$
1-\alpha\leq \mathbb{P}(T_{n+1} \in \widehat{C}_n(X_{n+1})) \leq 1-\alpha + \frac{1}{|\mathcal{I}_{ca}|+1},
$$
where $\mathcal{I}_{ca}$ denotes the number of calibration samples.
\end{lemma}

\textbf{Remark.} One may wonder why not use $S(t)$ to return the confidence band in Cox-MLP. 
People usually do so in practice, but the confidence band has no theoretical guarantee in Cox-MLP.
To derive the theoretical guarantee under $S(t)$, one needs to show the convergence of the predictor $g(X_i)$.
However, it cannot be proved unless the generalization guarantee of neural networks is obtained.

\section{Confidence band for the Survival Time}
In this section, we derive the confidence band for the survival time. 
We start by analyzing the basic properties and the critical ideas before proposing the algorithm.

\begin{algorithm*}[thbp]
\caption{WCCI: Weighted Conformal Censoring Inference} 
\label{alg1: CQSA1} 
\begin{algorithmic}[1] 
\REQUIRE Level $\alpha$
\REQUIRE data $\mathcal{Z} = (X_i, Y_i, \Delta_i)_{i \in \mathcal{I}}$, testing point $X^\prime$
\REQUIRE function $\hat{w}(x; \mathcal{D})$ to fit the weight function at $x$ using $\mathcal{D}$ as data

1. Randomly split $\mathcal{Z}$ into a training fold $\mathcal{Z}_{tr} \triangleq (X_i, Y_i, \Delta_i)_{i \in \mathcal{I}_{tr}}$ and a calibration fold $\mathcal{Z}_{ca} \triangleq (X_i, Y_i, \Delta_i)_{i \in \mathcal{I}_{ca}}$

2. Use $\mathcal{Z}_{tr}$ to train $\hat{g}(\cdot)$ to estimate the predictor function

3. For each $i \in \mathcal{I}_{ca}$ with $\Delta_i = 1$, compute the non-conformity score 
$V_i = \log (\sum_{k \in R(T_i) \cap \mathcal{I}_{tr}} \exp [\hat{g}(X_k) - \hat{g}(X_i)])$

4. For each $i \in \mathcal{I}_{ca}$ with $\Delta_i = 1$, compute the weight $W_i = \hat{w}(X_i; \mathcal{Z}_{tr})$

5. Calculate the normalized weights $\hat{p}_i = \frac{W_i}{\sum_{i \in \mathcal{I}_{ca}}{W_i}+\hat{w}(X^\prime; \mathcal{Z}_{tr})}$ and $\hat{p}_\infty = \frac{\hat{w}(X^\prime; \mathcal{Z}_{tr})}{\sum_{i \in \mathcal{I}_{ca}}{W_i}+\hat{w}(X^\prime; \mathcal{Z}_{tr})}$

6. Calculate the $(1-\alpha)$-th quantile $Q_{1-\alpha}$ of the distribution $\sum_{i \in \mathcal{I}_{ca}} \hat{p}_i \delta_{V_i} + \hat{p}_\infty {\delta}_{\infty}$

7. Calculate $T^u(X^\prime)$ as the smallest value such that its conformity score $V'$ (dependent on $T^u(X^\prime)$) is larger than $Q_{1-\alpha}$

\ENSURE $\hat{C}^1(X^\prime) = [0, T^u(X^\prime)]$.

\end{algorithmic} 
\end{algorithm*}

\begin{figure}
    \centering
    \includegraphics[width=0.96\linewidth]{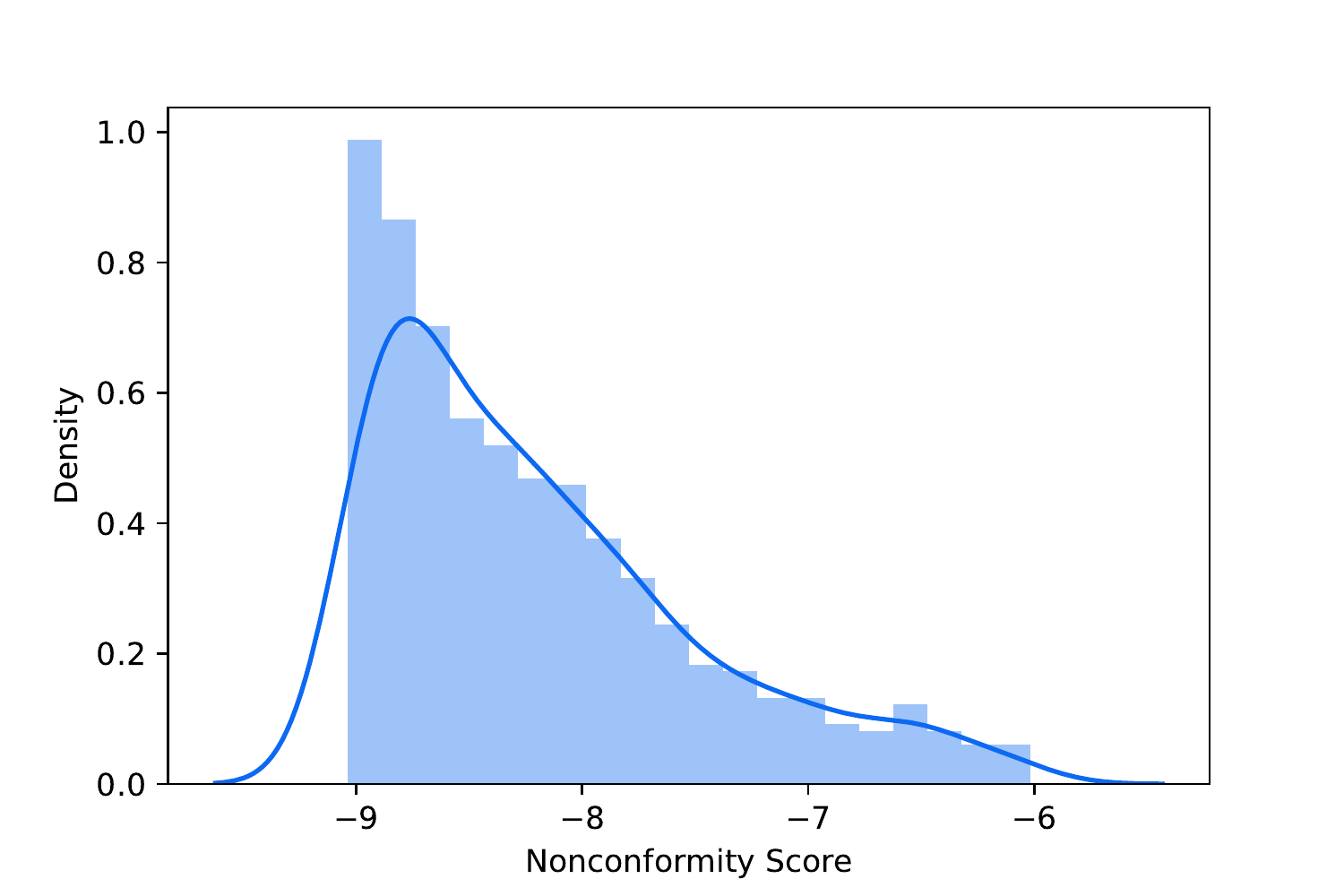}
     \caption{\textbf{Non-conformity score distribution.} We show the distribution of the non-conformity score\protect\footnotemark[2], which has a well-shaped single peak (more stable, see appendix for more details). The dark blue line is the KDE approximation.}
    \label{fig: non-con score}
\end{figure}

{\textbf{The non-conformity score.}}
In traditional conformal inference, the most commonly used non-conformity score is the absolute form $V_i = |T_i-\hat{T_i}|$. 
That is because we can directly derive the confidence band of $T_i$ based on the band of the non-conformity score $V_i$.
Unfortunately, it is hard to compute $\hat{T_i}$ in Cox-based models since they output the survival hazard function instead of the survival time, which requires a modification of the non-conformity score.

Inspired by the standard Cox regression, we introduce a non-conformity score based on partial likelihood. 
Specifically, we use the sample partial log-likelihood as the non-conformity score (shown in Equation~\ref{Eqn:conformalScore}).
Compared to the absolute non-conformity score, the newly proposed non-conformity score $V_i = V_g(X_i, T_i)$ do not need to calculate $\hat{T}_i$ explicitly. 
Figure~\ref{fig: non-con score} illustrates the distribution of the non-conformity score using a simulation dataset.
\begin{equation}
\label{Eqn:conformalScore}
V_i  = \log \left(\sum_{k \in R(T_i)} \exp [g(X_k) - g(X_i)]\right).
\end{equation} 

\textbf{The incomplete data.}
Compared to the i.i.d.~(independent and identically distributed) data, censored data is incomplete.
Notice that we can only calculate the non-conformity scores of the uncensored data in Equation~\ref{Eqn:conformalScore} since we cannot obtain the exact value of $T_i$ of censored data.
As a result, the returned confidence band is guaranteed under uncensored distribution $\mathcal{P}_{\xv|\iv=1}$.
However, as mentioned before, there might be a distribution shift between censored data and uncensored data, leading to the requirement of \emph{weighted conformal inference} \citep{tibshirani2019conformal}.

\footnotetext[2]{The simulation dataset is from \citet{kvamme2019time}}

In weighted conformal inference, one needs to calculate the weight $w$ based on $w(x)$, where
\begin{equation}
    \label{Eqn:weight}
    w(x) = \frac{\rm d \gP_X(x)}{\rm d \gP_{X|\Delta=1} (x)} = \frac{\mathbb{P}(T=1)}{\mathbb{P}(T=1\ | \ X=x )}.
\end{equation}

Intuitively, $w(x)$ helps transfer the guarantee over $\mathcal{P}_{\xv|\iv=1}$ to $\mathcal{P}_{\xv}$.
In the following of the paper, we use the regularized weight $p_i$ in the algorithm, namely
\begin{equation*}
    p_i = \frac{w(X_i)}{\sum_{i\in \mathcal{I}_{ca}} w(X_i) + w(X^\prime)}, 
\end{equation*}
where we denote $X_i$ as the samples from the calibration fold and $X^\prime$ as the testing point.

\textbf{Exchangeablility.}
In conformal inference, we assume that the non-conformity score satisfies exchangeablility (See Assumption~\ref{Exchangeability}).
However, as shown in Equation~\ref{Eqn:conformalScore}, there is a summation term in the non-conformity score, where we need to sum up all the samples at risk at time $T_j-$. 
Unfortunately, it breaks the exchangeablility when we use the at-risk samples in the calibration fold.
As an alternative, we use the at-risk samples in the training fold.
The non-conformity scores in the calibration fold $V_i,\ i\in[n]$ then satisfies exchangeablility given the training fold (See Equation~\ref{eqn: indenpendent}). 
\begin{equation}
\label{eqn: indenpendent}
(V_1, \dots, V_n\ |\ \gD_{tr}) \overset{d}{=} (V_{\pi(1)}, \dots, V_{\pi(n)}\ |\ \gD_{tr}),
\end{equation}
with arbitrary permutation $\pi$.

\textbf{The reconstruction of confidence band.}
Based on the non-conformity score proposed in Equation~\ref{Eqn:conformalScore}, we can reconstruct the confidence band for the survival time.
We remark that we calculate the one-sided band $[0, T^u(X^\prime)]$ although the two-sided band directly follows.
For a new sample $X^\prime$, we first calculate the $(1-\alpha)$ quantile of (weighted) distribution of the non-conformity score, denoted as $Q_{1-\alpha}$.
We then calculate the smallest $T^u(X^\prime)$ that makes its non-conformity score $V^\prime=V_{\hat{g}}(X^\prime, T)$ larger than $Q_{1-\alpha}$, formally,
\begin{equation*}
    \begin{split}
  Q_{1-\alpha} &=  \operatorname{Quantile}\left(1-\alpha, \sum_{i \in \mathcal{I}_{ca}} {p}_i V_i\right) \\
  T^u(X^\prime) &= \inf \left\{ T: V_{\hat{g}}(X^\prime, T) \geq Q_{1-\alpha} \right\}.     
    \end{split}
\end{equation*}

\textbf{Algorithm.}
Based on the discussions above, we conclude our algorithm in Algorithm~\ref{alg1: CQSA1}. 
In training process, we use training fold to train the estimated predictor $\hat{g}(\xv)$ (defined in Equation~\ref{Eqn: separable}) and calculate the weight function (defined in Equation~\ref{Eqn:weight}).
We then use calibration fold to calculate the non-conformity score.
We finally construct confidence band $[0, T^u(\xv^\prime)]$ for a given testing point $\xv^\prime$.

\textbf{Robustness on $\hat{w}$.}
The weighted conformal inference has guaranteed coverage under the true weight $w(x)$ as shown in Lemma~\ref{WCI}.
However, we can only obtain its estimator $\hat{w}(x)$ in practice.
In the following Theorem~\ref{Thm:robustness}, we prove how the estimation influences the coverage.
Note that as $\lim_{|\mathcal{Z}_{tr}| \to \infty} |\hat{w}(x) - {w}(x)| \to 0$, the coverage can be provably larger than $1-\alpha$.

\begin{restatable}[Provable Guarantee]{thm}{robustness}
\label{Thm:robustness}
Let $\hat{w}(x)$ be an estimate of the weight $w(x)$. 
Assume that $\mathbb{E}[\hat{w}(X)|\mathcal{Z}_{tr}] = 1$ and $\mathbb{E}[{w}(X)] = 1$.
Denote $\hat{C}_n^1(x)$ as the output band of Algorithm~\ref{alg1: CQSA1} with $n$ calibration samples, then for a new data $X^\prime$, its corresponding survival time satisfies
\begin{equation*} 
\lim_{n\to\infty}\mathbb{P}\left( T^\prime \in \hat{C}_n^1(X^\prime) \right) \geq 1-\alpha - \frac{1}{2} \mathbb{E}|\hat{w}(X) - {w}(X)|,
\end{equation*}
where the probability $\mathbb{P}$ on the left hand side is taken over ${(X^\prime, T^\prime)\sim \mathcal{P}_X \times \mathcal{P}_{T|X}}$, and all the expectation operators $\mathbb{E}$ are taken over $X \sim \mathcal{P}_{\xv|\iv=1}$.
\end{restatable}


Theorem~\ref{Thm:robustness} proves the lower bound for WCCI.
However, it is insufficient to derive the upper bound under the weighted conformal inference framework.
To derive the upper bound, we propose the algorithm T-SCI in the next section.

\section{T-SCI: Improved Estimation}

\begin{algorithm*} 
\caption{T-SCI: Two-Stage Conformal Inference} 
\label{alg1: New version} 
\begin{algorithmic}[1] 
\REQUIRE Level $\alpha$
\REQUIRE additional data $\mathcal{Z}_{ca2} = (X_i, Y_i, \Delta_i)_{i \in \mathcal{I}_{ca2}}$, testing point $X^\prime$
\REQUIRE First-stage band $[\hat{q}_{\alpha_{lo}}(X^\prime; \mathcal{Z}_{tr}, \mathcal{Z}_{ca1}), \hat{q}_{\alpha_{hi}}(X^\prime; \mathcal{Z}_{tr}, \mathcal{Z}_{ca1})]$ output from Algorithm~\ref{alg1: CQSA1} \protect\footnotemark[3] 
\REQUIRE function $\hat{w}(x; \mathcal{D})$ to fit the weight function at $x$ using $\mathcal{D}$ as data

\STATE For each $i \in \mathcal{I}_{ca2}$ with $\Delta_i = 1$, compute the non-conformity score $V_i = \max \{ \hat{q}_{\alpha_{lo}}(X_i; \mathcal{Z}_{tr}) - T_i, T_i - \hat{q}_{\alpha_{hi}}(X_i; \mathcal{Z}_{tr}) \}$

\STATE For each $i \in \mathcal{I}_{ca2}$ with $\Delta_i = 1$, compute the weight $W_i = \hat{w}(X_i; \mathcal{Z}_{tr})$

\STATE Calculate the normalized weights $\hat{p}_i = \frac{W_i}{\sum_{i \in \mathcal{I}_{ca}}{W_i}+\hat{w}(X^\prime; \mathcal{Z}_{tr})}$ and $\hat{p}_\infty = \frac{\hat{w}(X^\prime; \mathcal{Z}_{tr})}{\sum_{i \in \mathcal{I}_{ca}}{W_i}+\hat{w}(X^\prime; \mathcal{Z}_{tr})}$



\STATE Calculate $\eta$ as the $(1-\alpha)$-th quantile of the distribution $\sum_{i \in \mathcal{I}_{ca}} \hat{p}_i \delta_{V_i} + \hat{p}_\infty {\delta}_{\infty}$

\ENSURE $\hat{C}^2(X^\prime) = [\hat{q}_{\alpha_{lo}}(X^\prime; \mathcal{Z}_{tr}, \mathcal{Z}_{ca1}) - \eta, \hat{q}_{\alpha_{hi}}(X^\prime; \mathcal{Z}_{tr}, \mathcal{Z}_{ca1}) +\eta]$.

\end{algorithmic} 
\end{algorithm*}

In the previous section, we propose WCCI, which has lower bound coverage guarantees.
However, for better data efficiency, we want not only lower bound but also upper bounds.
To reach the goal, we propose a two-stage algorithm T-SCI, which returns nearly perfect coverage.

The intuition is from Lemma~\ref{QCI}, stating that the quantile conformal inference (QCI) returns a nearly perfect guarantee.
Inspired by Lemma~\ref{QCI}, we first use Algorithm~\ref{alg1: CQSA1} to return a temporal confidence band and then apply Quantile Conformal Inference to modify this band.
We summarize the whole algorithm in Algorithm~\ref{alg1: New version}.

\textbf{Remark 1.} To apply T-SCI in practice, we split the dataset into $\mathcal{Z}_{tr}$, $\mathcal{Z}_{ca1}$ and $\mathcal{Z}_{ca2}$. We first apply Algorithm~\ref{alg1: CQSA1} with $\mathcal{Z}_{tr}$ and $\mathcal{Z}_{ca1}$ and return a confidence band.
We then do calibration in Algorithm~\ref{alg1: New version} with $\mathcal{Z}_{ca2}$.

\textbf{Remark 2.} In Algorithm~\ref{alg1: New version}, we modify the conformal score to be the absolute form again for clarity, since we have already derived an interval estimator of $T$ in Algorithm~\ref{alg1: CQSA1}.

\footnotetext[3]{We use the two-sided band here for generality, and the one-sided band directly follows.}

Note that there are two conformal inference procedures in Algorithm~\ref{alg1: New version}.
When the context is clear, the weight function $w(x)$ and the non-conformity score $V_i$ refers to those in the second conformal inference. 
Before stating the theorem, we denote $H(X)$ to measure how well the quantile estimators $\hat{q}_{\alpha_{lo}}(X), \hat{q}_{\alpha_{hi}}(X)$ are.
\begin{equation}
\label{eqn: hx}
    H(X) = \max\{ |\hat{q}_{\alpha_{lo}}(X) - {q}_{\alpha_{lo}}(X)|, |\hat{q}_{\alpha_{hi}}(X) - {q}_{\alpha_{hi}}(X)| \}
\end{equation}
We next show in Theorem~\ref{thm: double robust} that Algorithm~\ref{alg1: New version} returns a guaranteed coverage either the weight or the temporal confidence band is estimated well.

\begin{restatable}[Lower Bound]{thm}{double}
\label{thm: double robust}
Let $\hat{w}(x)$ be an estimate of the weight $w(x)$,  $\hat{q}_{\alpha_{lo}}(x), \hat{q}_{\alpha_{hi}}(x)$ be the quantile estimator returned by WCCI, and $H(X)$ be defined as Equation~\ref{eqn: hx}.  
Assume that $\mathbb{E}[\hat{w}(X)|\mathcal{Z}_{tr}] = 1$ and $\mathbb{E}[{w}(X)] = 1$, where all the expectation operators $\mathbb{E}$ are taken over $X \sim \mathcal{P}_{\xv|\iv=1}$.
Denote $\hat{C}_n^2(x)$ as the output band of Algorithm~\ref{alg1: New version} with $n$ calibration samples, and denote $X^\prime$ as the testing point.

From the weight perspective, under assumptions (A1):\\
\textbf{\emph{A1.}} $\mathbb{E}_{\xv|\iv=1}|\hat{w}(X) - {w}(X)| \leq M_1,$\\
we have:
\begin{equation*} 
\lim_{n\to\infty}\mathbb{P}\left( T^\prime \in \hat{C}_n^2(X^\prime) \right) \geq 1-\alpha - \frac{1}{2} M_1.
\end{equation*}

From the quantile perspective, under assumptions (B1-B3):\\
\textbf{\emph{B1.}} $H(X) \leq M_2$ a.s. w.r.t. $X$;\\
\textbf{\emph{B2.}} There exists $\delta>0$ such that $\mathbb{E} \hat{w}(X)^{1+\delta} < \infty$;\\
\textbf{\emph{B3.}} There exists $\gamma, b_1, b_2>0$ such that $\mathbb{P}(T=t|X=x)\in [b_1, b_2]$ uniformly over all (x,t) with $t \in [{q}_{\alpha_{lo}}(x)-2M_2-2\gamma, {q}_{\alpha_{lo}}(x)+2M_2+2\gamma] \cup  [{q}_{\alpha_{hi}}(x)-2M_2-2\gamma, {q}_{\alpha_{hi}}(x)+2M_2+2\gamma] $,\\
we have:
\begin{small}
\begin{equation*}
\lim_{n\to\infty}\mathbb{P}\left( T^\prime \in \hat{C}_n^2(X^\prime) \right)
\geq  1 - \alpha -b_2 (2 M_2 + \gamma) -  \frac{16 M_2}{(M_2+\gamma)^2 b_1}.    
\end{equation*}
\end{small}
\end{restatable}

Theorem~\ref{thm: double robust} demonstrates that when $M_1 \to 0$ \emph{or} $M_2, \gamma \to 0$ as $|\mathcal{Z}_{tr}| \to 0$, the confidence band has guaranteed coverage with lower bound $1-\alpha$.
Compared to Theorem~\ref{Thm:robustness}, Theorem~\ref{thm: double robust} is doubly robust since the coverage is guaranteed when either (A1) or (B1-B3) holds.
We next prove the upper bound in Theorem~\ref{thm: upperbound}.

\begin{restatable}[Upper Bound]{thm}{upperBound}
\label{thm: upperbound}
Let $\hat{w}(x)$ be an estimate of the weight $w(x)$,  $\hat{q}_{\alpha_{lo}}(x), \hat{q}_{\alpha_{hi}}(x)$ be the quantile estimator returned by WCCI, and $H(X)$ be defined as Equation~\ref{eqn: hx}.  
Assume that $\mathbb{E}[\hat{w}(X)|\mathcal{Z}_{tr}] = 1$ and $\mathbb{E}[{w}(X)] = 1$, where all the expectation operators $\mathbb{E}$ are taken over $X \sim \mathcal{P}_{\xv|\iv=1}$.
Let $F_V \triangleq \sum_{i \in \mathcal{I}_{ca}} {p}_i \delta_{V_i} + {p}_\infty {\delta}_{\infty}$ be CDF, and assume $V_i$ has no ties.
Denote $\hat{C}_n^2(x)$ as the output band of Algorithm~\ref{alg1: New version} with $n$ calibration samples, and $X^\prime$ as the testing point.\\
Under assumptions (C1-C4):\\
\textbf{\emph{C1.}} $\forall \mathcal{S} \subset \mathcal{I}_{ca2}$, $|\sum_{i \in \mathcal{S}} (w(X_i) - \hat{w}(X_i))| \leq M_1^\prime$;\\
\textbf{\emph{C2.}} $H(X) \leq M_2^\prime$ a.s. w.r.t. $X$;\\
\textbf{\emph{C3.}} $F_V(t+L) - F_V(t) \geq K L$ for all $t, L$;\\
\textbf{\emph{C4.}} there exists $b_1, b_2>0$ such that $\mathbb{P}(T=t|X=x)\in [b_1, b_2]$ uniformly over all (x,t) with $t \in [{q}_{\alpha_{lo}}(x)-r, {q}_{\alpha_{lo}}(x)+r] \cup  [{q}_{\alpha_{hi}}(x)-r, {q}_{\alpha_{hi}}(x)+r] $, where $r = 2M_2^\prime+2M_1^\prime/K$
We have
$$
\lim_{n\to\infty}\mathbb{P}\left( T^\prime \in \hat{C}_n^2(X^\prime) \right) \leq 1 - \alpha + b_2 (2 M_2^\prime + \frac{M_1^\prime}{K}).
$$
\end{restatable}

Theorem~\ref{thm: upperbound} demonstrates that when the weight function $\hat{w}(x)$ and the quantile function $\hat{q}_{\alpha_{lo}}(x), \hat{q}_{\alpha_{hi}}(x)$ are estimated well (C1-C2), the returned coverage T-SCI has a lower bound guarantee.
Combining Theorem~\ref{thm: double robust} and Theorem~\ref{thm: upperbound} leads to a ``nearly perfect'' guaranteed coverage for T-SCI.



\section{Experiments}
This section aims at verifying some key arguments: 
(1) T-SCI returns valid coverage with small length interval; 
(2) Weight plays an essential role in the algorithm; 
(3) Censoring is more challenging than uncensoring settings.
The results support these arguments both in synthetic data and real-world data, see Table~\ref{tab:rr_res}.

\begin{table*}[t]
    \centering
    \caption{Performance of Different Models on RRNLNPH. }
    \begin{tabular}{ccccccccc}
    \hline 
        Method & \multicolumn{2}{c}{Total} & \multicolumn{2}{c}{Censored} & \multicolumn{2}{c}{Uncensored} & \multicolumn{2}{c}{Interval Length}\\
        \cline{2-9}
         & Mean & Std.  & Mean & Std.  & Mean & Std.  & Mean & Std.\\
         \hline\hline 
         Cox Reg. & 0.832 & 0.008 & / & / & / & / & 22.08 & 0.23\\
         Random Survival Forest \cite{ishwaran2008random} & 0.948 & 0.006 & / &/ &/  & /& 16.39&0.22\\
         Nnet-Survival \cite{gensheimer2019scalable}  & 0.834 & 0.007 & 0.560 &0.016 & 0.982 & 0.005 & 20.11 &0.37 \\
         MTLR \cite{yu2011learning}  & 0.830 & 0.008 & 0.554 & 0.017 & 0.980 & 0.003 & 19.85 & 0.34\\
        CoxPH \citep{katzman2018deepsurv} & 0.829 & 0.008 &0.554 &0.016 & 0.978 &0.004 & 19,65 & 0.21\\
        CoxCC \citep{kvamme2019time} & 0.830 & 0.008 & 0.556& 0.016& 0.975 & 0.003 & 20.17 & 0.24\\
        CoxPH+WCCI & 0.912 & 0.03 & 0.854 & 0.047 & 0.949 & 0.028& 21.59 & 0.90\\
        CoxPH+T-SCI & 0.974 & 0.009 & 0.947 & 0.018 & 0.994 & 0.005 & 29.85 & 0.68\\
        CoxCC+WCCI & 0.919 & 0.03 & 0.862 & 0.043 & 0.955 & 0.026& 21.55 & 1.27\\ 
        CoxCC+T-SCI & 0.974 & 0.009 & 0.946 & 0.017 & 0.995 & 0.004 & 29.62 & 0.57\\
        CoxPH+WCCI(unweigted) & 0.907 & 0.020 & 0.830 & 0.049 & 0.949 & 0.006 & 22.06 & 1.27\\
        CoxPH+T-SCI(unweighted) & 0.950 & 0.018 & 0.875 & 0.049 & 0.990 & 0.012 & 27.72 & 3.13\\
        CoxCC+WCCI(unweigted) & 0.941 & 0.029  & 0.815 & 0.029 & 0.948 & 0.009 & 22.57 & 0.70\\
        CoxCC+T-SCI(unweighted) & 0.955 & 0.020 & 0.877 & 0.048 & 0.992 & 0.007 & 28.92 & 1.04\\
        Kernel \citep{chen2020deep} & 0.951 & 0.024 & 0.858 & 0.093 & 0.993 &  0.014 & 51.63 & 32.92\\
        \hline 
    \end{tabular}
    \label{tab:rr_res}
    \vspace{-10pt}
\end{table*}

\subsection{Setup}
\textbf{Datasets and Environment.}
We conduct extensive experiments to test the efficiency of the algorithm. 
We use one synthetic dataset {RRNLNPH} (from \citet{kvamme2019time}) and two real-world datasets, METABRIC and SUPPORT (See supplementary materials for more details). 
For each dataset, we test several baseline algorithms along with our proposed algorithms.
In each run, 80\% data are randomly sampled as the training data, and the two halves of the rest are randomly split as calibration data and test data.
We run the experiments 100 times for each algorithm. 
Moreover, we collect the results under different $\alpha$s. 

\textbf{Algorithms.}
We choose the linear Cox regression (labeled as Cox Reg.) as a baseline.
Besides, we conduct CoxPH \citep{katzman2018deepsurv} and CoxCC \citep{kvamme2019time} (both belong to Cox-MLP) using $S(t)$ to return confidence band although they do not contain a theoretical guarantee.
We also choose a kernel-based non-Cox method, \citep{chen2020deep} (labeled as Kernel).
Besides, we conduct RSF \citep{ishwaran2008random} (labeled as Random Survival Forest), Nnet-Survival \citep{gensheimer2019scalable} (labeled as Nnet-Survival), and MTLR \citep{yu2011learning} (labeled as MTLR) as benchmarks.
We emphasize that these methods are different approaches since our method is Cox-based. 

We integrate our proposed algorithms WCCI and T-SCI with CoxPH and CoxCC, labeled as CoxPH/CoxCC + WCCI/T-SCI.
Besides, 
to ensure that the weights in WCCI and T-SCI are helpful, we test the \emph{unweighted version} of WCCI and T-SCI, where we set all weights to 1 in WCCI and T-SCI.
We summarize all the experimental results under significance level $\alpha = 95\%$ in Table~\ref{tab:rr_res}.
Ideally, a perfect method returns coverage slightly larger to $95\%$ with small interval length while being balanced in censored and uncensored data.

\textbf{Metrics.}
In the synthetic dataset, our core metric is \emph{empirical coverage} (EC), defined as the fraction of testing points whose survival time falls in the predicted confidence band.
Besides, we calculate the \emph{average interval length} of the returned band.
Given confidence level $\alpha$, a perfect confidence band is expected to return empirical coverage larger than $1-\alpha$ with a small interval length.
In the real-world dataset, we use \emph{surrogate empirical coverage} (SEC) which is the upper bound of EC.
We defer its formal definition in the supplementary materials.

\begin{figure}[t]
    \centering
    \includegraphics[width=0.96\linewidth]{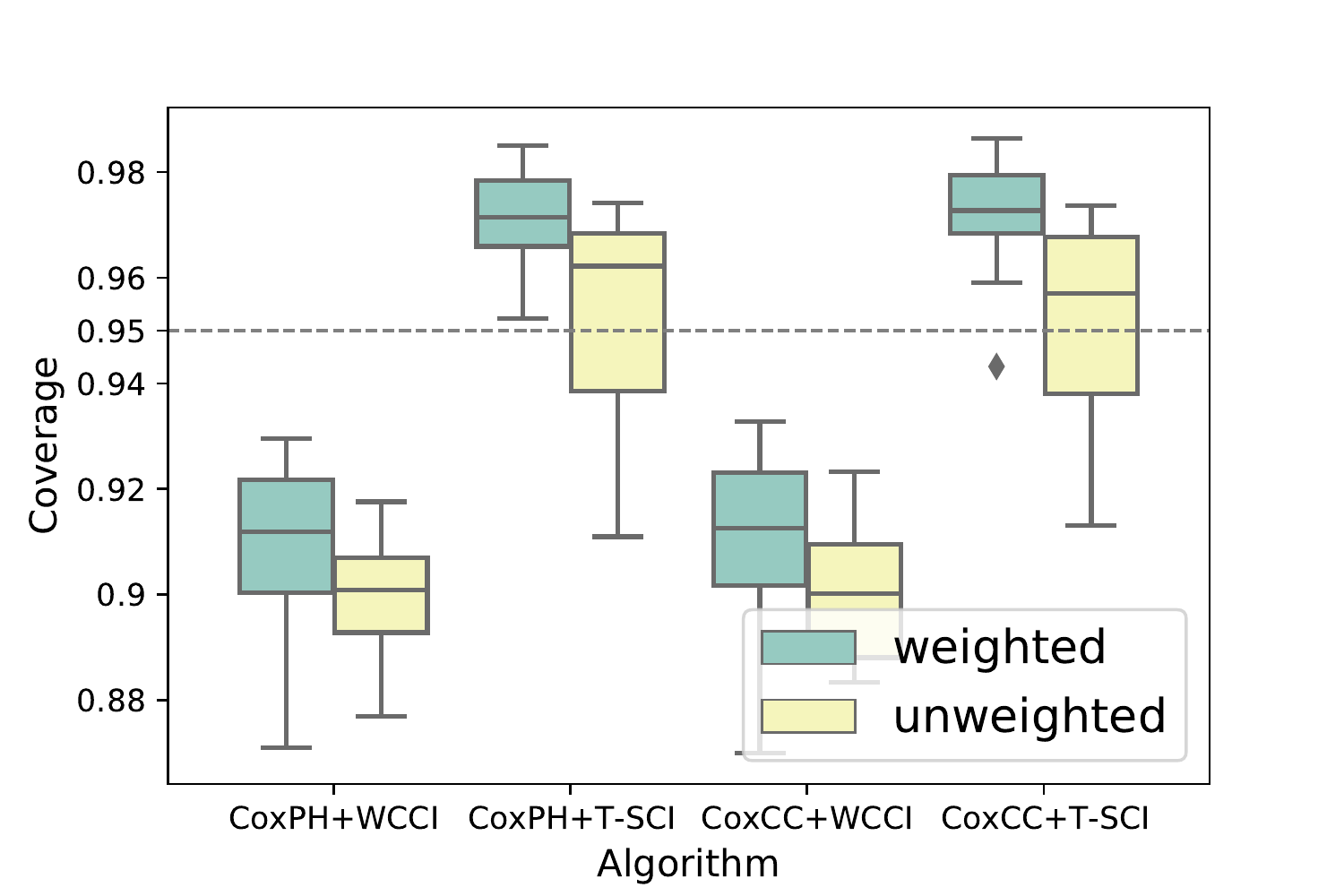}
    \caption{\textbf{Weight Rationality.} We compare the weighted version (green) with its corresponding unweighted version (yellow).
     The weighted versions contain less bias for WCCI (closer to the expected $95\%$) and less variance for T-SCI (shorter boxes).}
    \label{fig: weight}
    \vspace{-7pt}
\end{figure}

\begin{figure}[t]
    \centering
    \includegraphics[width=0.96\linewidth]{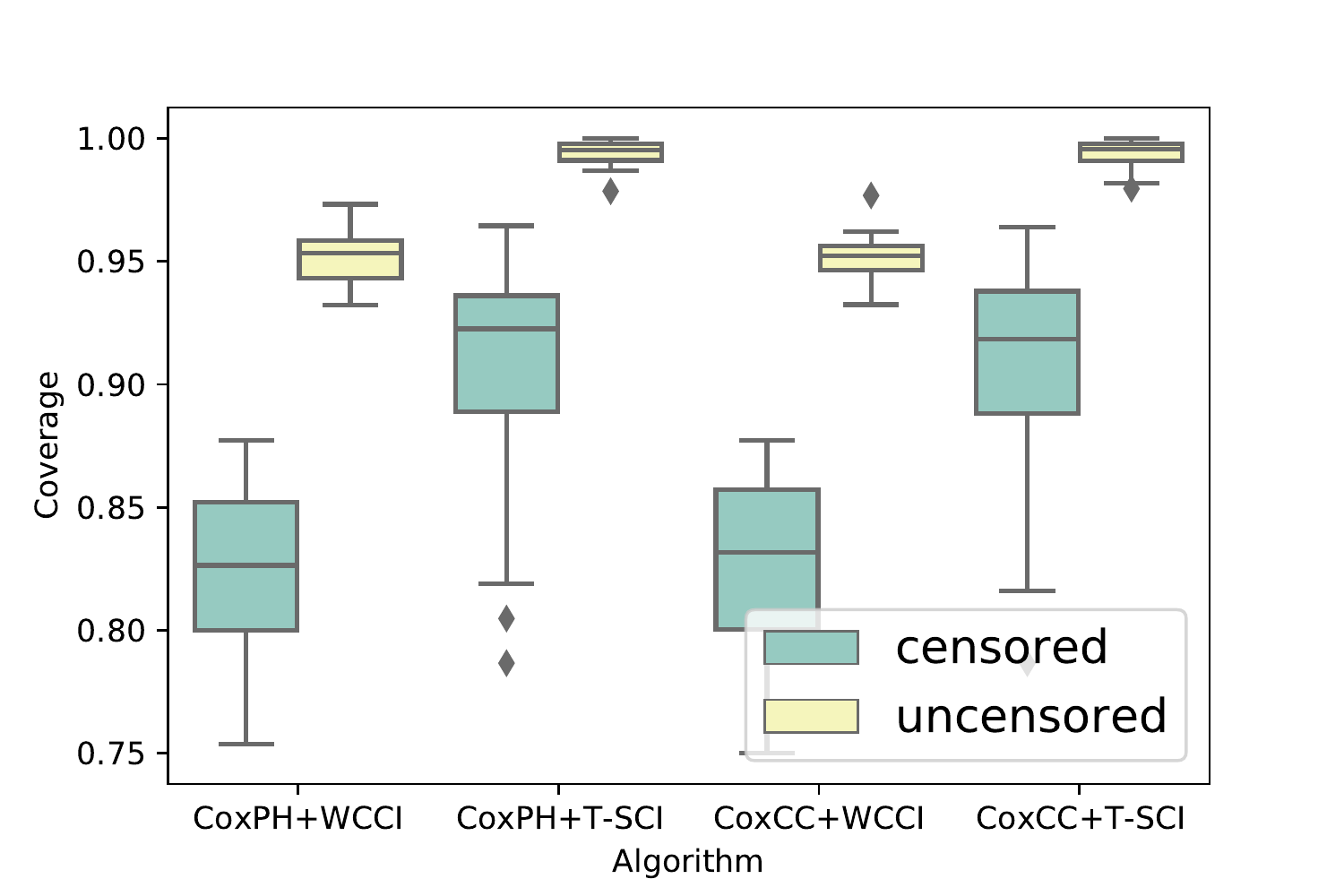}
    \caption{\textbf{Censoring comparison.} We compare model performances on censored (green) and uncensored (uncensored) data separately. All the algorithms show a larger coverage and less variance on censored data. }
    \label{fig: censor}
    \vspace{-7pt}
\end{figure}

\subsection{Analysis}
We summarize the experimental results on RRNLNPH in Table~\ref{tab:rr_res} and defer the results on SUPPORT and METABRIC to the supplementary materials due to space limitations.

\begin{figure*}[t]
\centering
\vspace{-15pt}
    \begin{subfigure}[htbp]{.49\textwidth}
    \setlength{\abovecaptionskip}{0cm}

    \includegraphics[width=8cm]{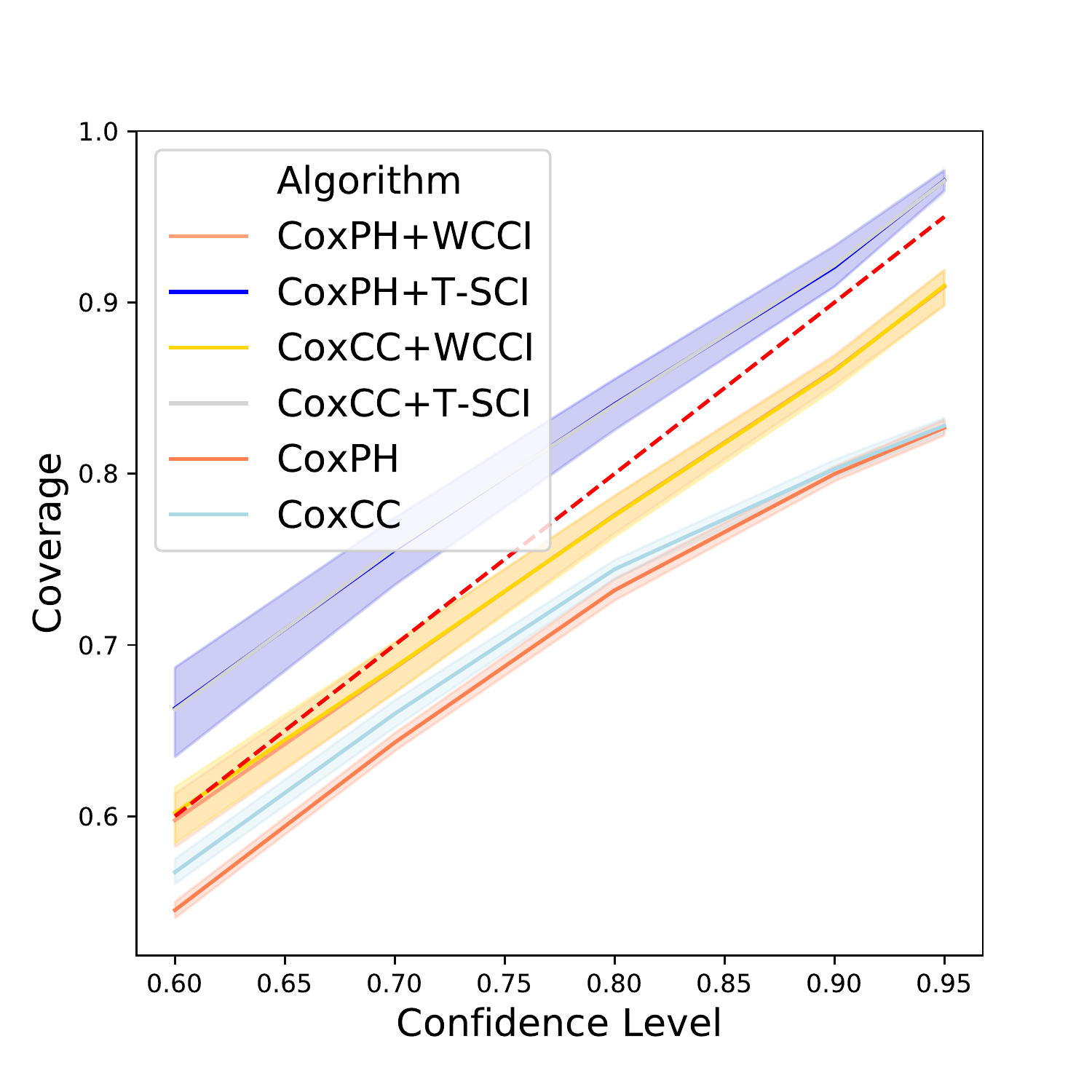}
    \caption{Coverage Comparison(weighted)}
    \end{subfigure}
    \begin{subfigure}[htbp]{.49\textwidth}
    \setlength{\abovecaptionskip}{0cm}

    \includegraphics[width=8cm]{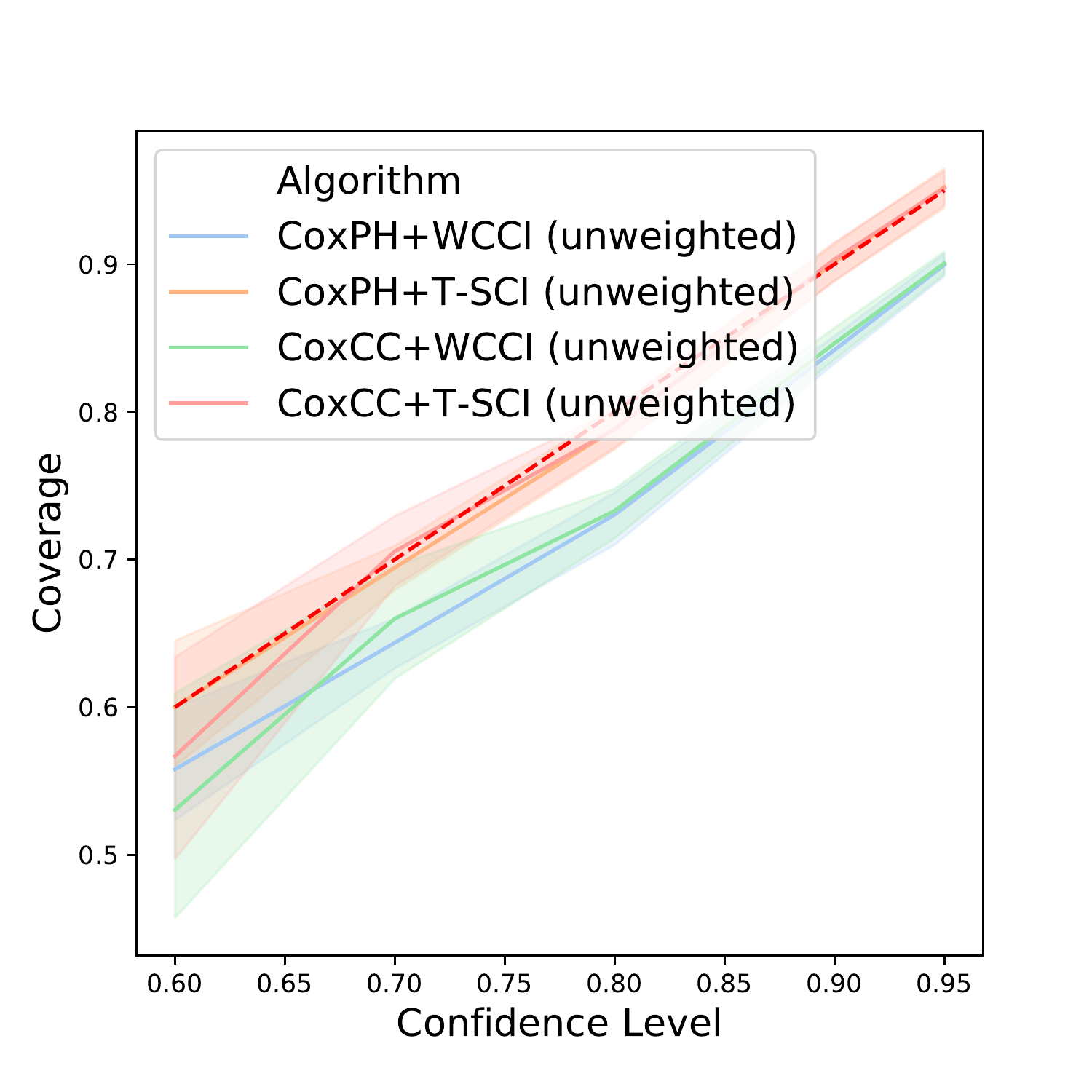}
    \caption{Coverage Comparison(unweighted)}
    \end{subfigure}
    \begin{subfigure}[htbp]{.49\textwidth}
    \setlength{\abovecaptionskip}{0cm}
    \includegraphics[width=8cm]{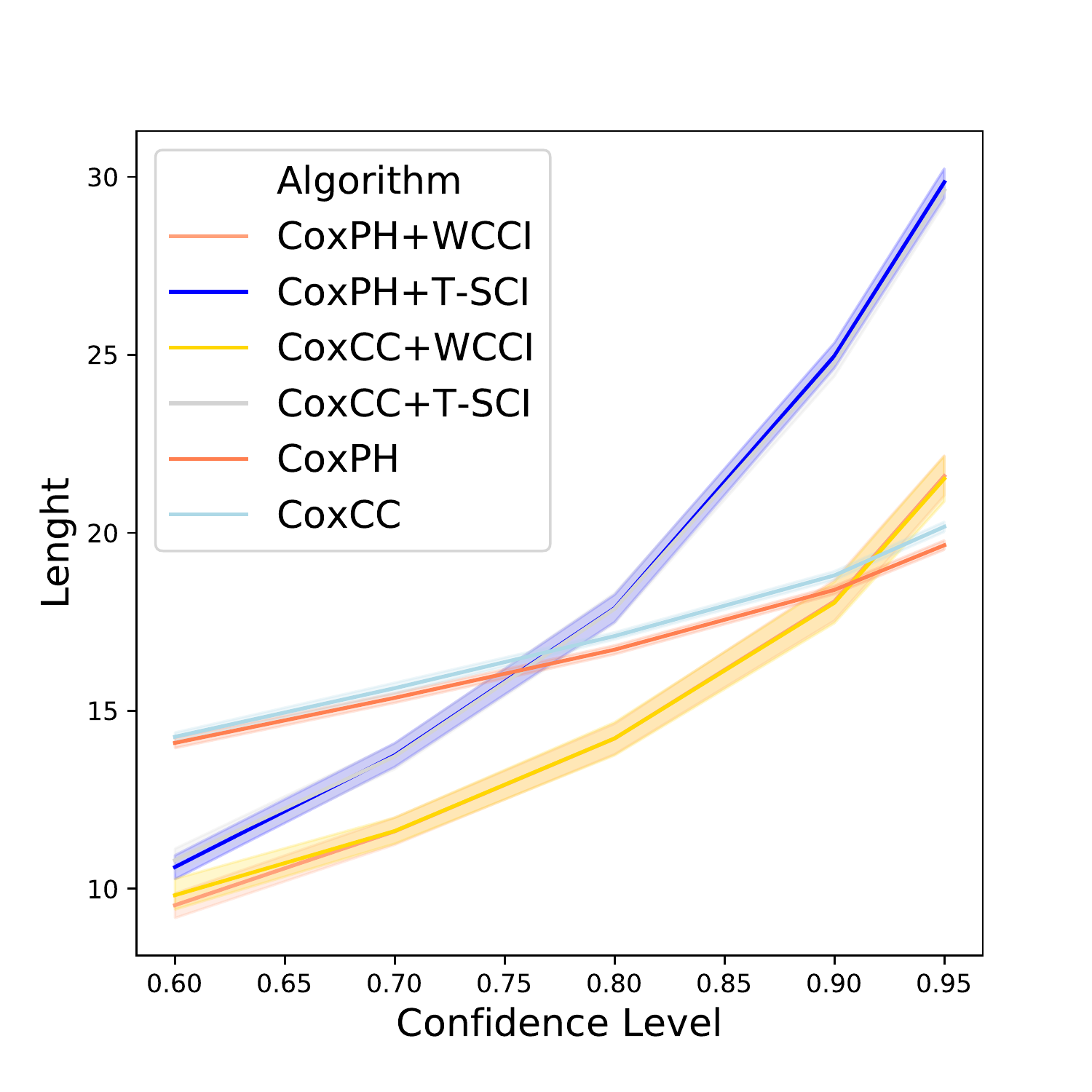}
    \caption{Interval Length Comparison(weighted)}
    \end{subfigure}
    \begin{subfigure}[htbp]{.49\textwidth}
    \setlength{\abovecaptionskip}{0cm}
    \includegraphics[width=8cm]{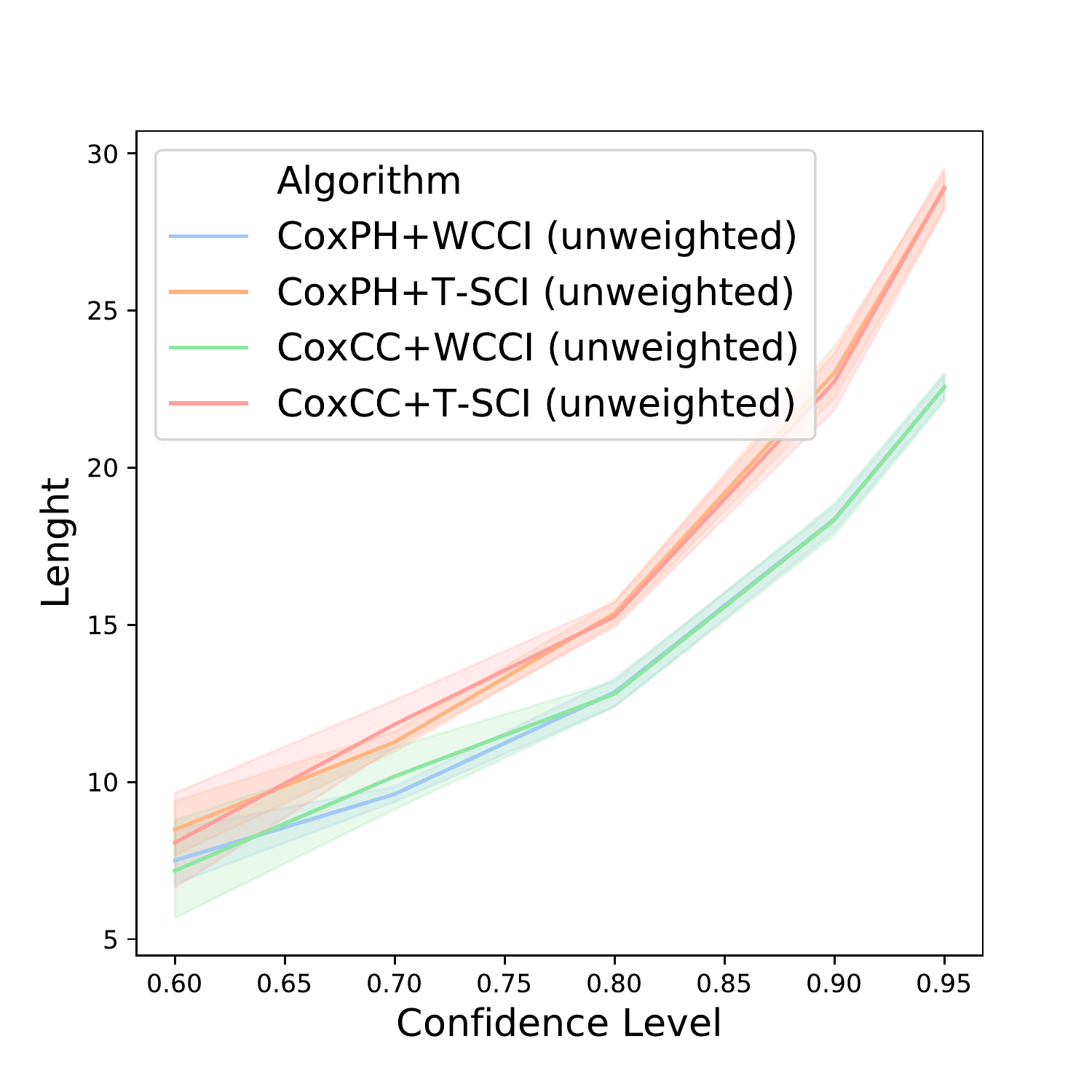}
    \caption{Interval Length Comparison(unweighted)}
    \end{subfigure}
    \caption{\textbf{Comparison under different confidence level ($1-\alpha$).} 
    CoxCC+T-SCI (purple) and CoxPH+T-SCI (grey)  returns guaranteed coverage under different confidence level without much increase of interval length (CoxCC+T-SCI and CoxPH+T-SCI overlap).}
    \label{fig: coveragelength}
\end{figure*}

\textbf{Coverage and Interval Length.}
Figure~\ref{fig: coveragelength} shows the empirical coverage and the interval length of the confidence band returned by algorithms under different confidence levels.
Ideally, the confidence band should have large coverage with a small interval length.
Notice that WCCI and T-SCI based algorithms outperform their original versions, showing that the proposed algorithms work well.
Furthermore, T-SCI is more conservative than WCCI, where T-SCI has larger coverage and interval length.
We further remark that T-SCI returns guaranteed coverage (larger than $1-\alpha$) under different confidence levels.

\textbf{The rationality of weight $w(x)$.}
We show the comparison between weighted and unweighted versions in Figure~\ref{fig: weight}. 
For WCCI algorithms, the weighted version has coverage closer to $1-\alpha$.
For T-SCI algorithms, the weighted version has lower variance. 
These results show that weighted versions outperform the unweighted versions, which validates the importance of weight $w(x)$. 

\textbf{The difficulty in censoring.}
Figure~\ref{fig: censor} shows the algorithms' performance on censored and uncensored data, respectively.
All the algorithms perform larger coverage and less variance on the uncensored data, meaning that censored data is more challenging to deal with than uncensored data.
Besides, we emphasize that although the unweighted versions may be closer to 95\% in some cases, they lack theoretical guarantees and are imbalanced, meaning that it performs pretty differently on censored and uncensored data.


\textbf{Analysis of Table~\ref{tab:rr_res}.}
We show RRNLNPH results in Table~\ref{tab:rr_res}, and results of SUPPORT and METABRIC perform similarly.
Firstly, notice that WCCI does not reach the expected coverage mainly due to the inaccurate estimation on the weight, while T-SCI reaches it due to milder requirements in Theorem~\ref{thm: double robust} (double robustness).
Secondly, notice that Cox Reg., CoxPH, CoxCC (they all lack theoretical guarantee) all fail to return the proper coverage.
Thirdly, unweighted versions all suffer from poor performances on censored data due to the lack of weight, showing that weight is vital in covariate shift.
Finally, we emphasize that the Kernel method suffers from large interval length and large variance despite the moderate coverage.
As a comparison, WCCI and T-SCI often perform more stably.

\section{Conclusion}
In this paper, we derive confidence band for Cox-based models.
We first introduce WCCI by proposing a new non-conformity score.
We then propose T-SCI, a two-stage conformal inference applying WCCI as input.
Theoretical analysis shows that T-SCI returns nearly perfect coverage, meaning both lower and upper bound guarantee.
We conduct extensive experiments on both synthetic data and real-world data to show the proposed algorithm's correctness.


\clearpage
\bibliography{example_paper}

\begin{thebibliography}{39}
\providecommand{\natexlab}[1]{#1}
\providecommand{\url}[1]{\texttt{#1}}
\expandafter\ifx\csname urlstyle\endcsname\relax
  \providecommand{\doi}[1]{doi: #1}\else
  \providecommand{\doi}{doi: \begingroup \urlstyle{rm}\Url}\fi

\bibitem[Akritas et~al.(1995)Akritas, Murphy, and Lavalley]{akritas1995theil}
Akritas, M.~G., Murphy, S.~A., and Lavalley, M.~P.
\newblock The theil-sen estimator with doubly censored data and applications to
  astronomy.
\newblock \emph{Journal of the American Statistical Association}, 90\penalty0
  (429):\penalty0 170--177, 1995.

\bibitem[Barber et~al.(2019{\natexlab{a}})Barber, Candes, Ramdas, and
  Tibshirani]{barber2019conformal}
Barber, R.~F., Candes, E.~J., Ramdas, A., and Tibshirani, R.~J.
\newblock Conformal prediction under covariate shift.
\newblock \emph{arXiv preprint arXiv:1904.06019}, 2019{\natexlab{a}}.

\bibitem[Barber et~al.(2019{\natexlab{b}})Barber, Candes, Ramdas, and
  Tibshirani]{barber2019limits}
Barber, R.~F., Candes, E.~J., Ramdas, A., and Tibshirani, R.~J.
\newblock The limits of distribution-free conditional predictive inference.
\newblock \emph{arXiv preprint arXiv:1903.04684}, 2019{\natexlab{b}}.

\bibitem[Bellotti \& Crook(2009)Bellotti and Crook]{bellotti2009credit}
Bellotti, T. and Crook, J.
\newblock Credit scoring with macroeconomic variables using survival analysis.
\newblock \emph{Journal of the Operational Research Society}, 60\penalty0
  (12):\penalty0 1699--1707, 2009.

\bibitem[Berrett et~al.(2020)Berrett, Wang, Barber, and
  Samworth]{berrett2020conditional}
Berrett, T.~B., Wang, Y., Barber, R.~F., and Samworth, R.~J.
\newblock The conditional permutation test for independence while controlling
  for confounders.
\newblock \emph{Journal of the Royal Statistical Society: Series B (Statistical
  Methodology)}, 82\penalty0 (1):\penalty0 175--197, 2020.

\bibitem[Bostr et~al.(2017)Bostr, Asker, Gurung, Karlsson, Lindgren,
  Papapetrou, et~al.]{bostr2017conformal}
Bostr, H., Asker, L., Gurung, R., Karlsson, I., Lindgren, T., Papapetrou, P.,
  et~al.
\newblock Conformal prediction using random survival forests.
\newblock In \emph{2017 16th IEEE International Conference on Machine Learning
  and Applications (ICMLA)}, pp.\  812--817. IEEE, 2017.

\bibitem[Bostr{\"o}m et~al.(2019)Bostr{\"o}m, Johansson, and
  Vesterberg]{bostrom2019predicting}
Bostr{\"o}m, H., Johansson, U., and Vesterberg, A.
\newblock Predicting with confidence from survival data.
\newblock In \emph{Conformal and Probabilistic Prediction and Applications},
  pp.\  123--141, 2019.

\bibitem[Breslow(1975)]{breslow1975analysis}
Breslow, N.~E.
\newblock Analysis of survival data under the proportional hazards model.
\newblock \emph{International Statistical Review/Revue Internationale de
  Statistique}, pp.\  45--57, 1975.

\bibitem[Cand{\`e}s et~al.(2021)Cand{\`e}s, Lei, and
  Ren]{candes2021conformalized}
Cand{\`e}s, E.~J., Lei, L., and Ren, Z.
\newblock Conformalized survival analysis.
\newblock \emph{arXiv preprint arXiv:2103.09763}, 2021.

\bibitem[Chen(2020)]{chen2020deep}
Chen, G.~H.
\newblock Deep kernel survival analysis and subject-specific survival time
  prediction intervals.
\newblock In \emph{Machine Learning for Healthcare Conference}, pp.\  537--565.
  PMLR, 2020.

\bibitem[Cox(1972)]{cox1972regression}
Cox, D.~R.
\newblock Regression models and life-tables.
\newblock \emph{Journal of the Royal Statistical Society: Series B
  (Methodological)}, 34\penalty0 (2):\penalty0 187--202, 1972.

\bibitem[Feldmann(2019)]{feldmann2019leo}
Feldmann, R.
\newblock Leo-py: Estimating likelihoods for correlated, censored, and
  uncertain data with given marginal distributions.
\newblock \emph{Astronomy and Computing}, 29:\penalty0 100331, 2019.

\bibitem[Gensheimer \& Narasimhan(2019)Gensheimer and
  Narasimhan]{gensheimer2019scalable}
Gensheimer, M.~F. and Narasimhan, B.
\newblock A scalable discrete-time survival model for neural networks.
\newblock \emph{PeerJ}, 7:\penalty0 e6257, 2019.

\bibitem[Ishwaran et~al.(2008)Ishwaran, Kogalur, Blackstone, Lauer,
  et~al.]{ishwaran2008random}
Ishwaran, H., Kogalur, U.~B., Blackstone, E.~H., Lauer, M.~S., et~al.
\newblock Random survival forests.
\newblock \emph{Annals of Applied Statistics}, 2\penalty0 (3):\penalty0
  841--860, 2008.

\bibitem[Kaplan \& Meier(1958)Kaplan and Meier]{kaplan1958nonparametric}
Kaplan, E.~L. and Meier, P.
\newblock Nonparametric estimation from incomplete observations.
\newblock \emph{Journal of the American statistical association}, 53\penalty0
  (282):\penalty0 457--481, 1958.

\bibitem[Katzman et~al.(2018)Katzman, Shaham, Cloninger, Bates, Jiang, and
  Kluger]{katzman2018deepsurv}
Katzman, J.~L., Shaham, U., Cloninger, A., Bates, J., Jiang, T., and Kluger, Y.
\newblock Deepsurv: personalized treatment recommender system using a cox
  proportional hazards deep neural network.
\newblock \emph{BMC medical research methodology}, 18\penalty0 (1):\penalty0
  1--12, 2018.

\bibitem[Klein \& Moeschberger(2006)Klein and Moeschberger]{klein2006survival}
Klein, J.~P. and Moeschberger, M.~L.
\newblock \emph{Survival analysis: techniques for censored and truncated data}.
\newblock Springer Science \& Business Media, 2006.

\bibitem[Kvamme et~al.(2019)Kvamme, Borgan, and Scheel]{kvamme2019time}
Kvamme, H., Borgan, {\O}., and Scheel, I.
\newblock Time-to-event prediction with neural networks and cox regression.
\newblock \emph{Journal of machine learning research}, 20\penalty0
  (129):\penalty0 1--30, 2019.

\bibitem[Lee et~al.(2018)Lee, Zame, Yoon, and van~der Schaar]{lee2018deephit}
Lee, C., Zame, W., Yoon, J., and van~der Schaar, M.
\newblock Deephit: A deep learning approach to survival analysis with competing
  risks.
\newblock In \emph{Proceedings of the AAAI Conference on Artificial
  Intelligence}, volume~32, 2018.

\bibitem[Lei \& Wasserman(2014)Lei and Wasserman]{lei2014distribution}
Lei, J. and Wasserman, L.
\newblock Distribution-free prediction bands for non-parametric regression.
\newblock \emph{Journal of the Royal Statistical Society: Series B: Statistical
  Methodology}, pp.\  71--96, 2014.

\bibitem[Lei et~al.(2013)Lei, Robins, and Wasserman]{lei2013distribution}
Lei, J., Robins, J., and Wasserman, L.
\newblock Distribution-free prediction sets.
\newblock \emph{Journal of the American Statistical Association}, 108\penalty0
  (501):\penalty0 278--287, 2013.

\bibitem[Lei et~al.(2018)Lei, G’Sell, Rinaldo, Tibshirani, and
  Wasserman]{lei2018distribution}
Lei, J., G’Sell, M., Rinaldo, A., Tibshirani, R.~J., and Wasserman, L.
\newblock Distribution-free predictive inference for regression.
\newblock \emph{Journal of the American Statistical Association}, 113\penalty0
  (523):\penalty0 1094--1111, 2018.

\bibitem[Lei \& Cand{\`e}s(2020)Lei and Cand{\`e}s]{lei2020conformal}
Lei, L. and Cand{\`e}s, E.~J.
\newblock Conformal inference of counterfactuals and individual treatment
  effects.
\newblock \emph{arXiv preprint arXiv:2006.06138}, 2020.

\bibitem[Li \& Bradic(2020)Li and Bradic]{li2020censored}
Li, A.~H. and Bradic, J.
\newblock Censored quantile regression forest.
\newblock In \emph{International Conference on Artificial Intelligence and
  Statistics}, pp.\  2109--2119. PMLR, 2020.

\bibitem[Nakagawa \& Freckleton(2008)Nakagawa and
  Freckleton]{nakagawa2008missing}
Nakagawa, S. and Freckleton, R.~P.
\newblock Missing inaction: the dangers of ignoring missing data.
\newblock \emph{Trends in ecology \& evolution}, 23\penalty0 (11):\penalty0
  592--596, 2008.

\bibitem[Nouretdinov et~al.(2011)Nouretdinov, Costafreda, Gammerman,
  Chervonenkis, Vovk, Vapnik, and Fu]{nouretdinov2011machine}
Nouretdinov, I., Costafreda, S.~G., Gammerman, A., Chervonenkis, A., Vovk, V.,
  Vapnik, V., and Fu, C.~H.
\newblock Machine learning classification with confidence: application of
  transductive conformal predictors to mri-based diagnostic and prognostic
  markers in depression.
\newblock \emph{Neuroimage}, 56\penalty0 (2):\penalty0 809--813, 2011.

\bibitem[Robins \& Finkelstein(2000)Robins and
  Finkelstein]{robins2000correcting}
Robins, J.~M. and Finkelstein, D.~M.
\newblock Correcting for noncompliance and dependent censoring in an aids
  clinical trial with inverse probability of censoring weighted (ipcw) log-rank
  tests.
\newblock \emph{Biometrics}, 56\penalty0 (3):\penalty0 779--788, 2000.

\bibitem[Romano et~al.(2019)Romano, Patterson, and
  Cand{\`e}s]{romano2019conformalized}
Romano, Y., Patterson, E., and Cand{\`e}s, E.~J.
\newblock Conformalized quantile regression.
\newblock \emph{arXiv preprint arXiv:1905.03222}, 2019.

\bibitem[Romano et~al.(2020)Romano, Sesia, and
  Cand{\`e}s]{romano2020classification}
Romano, Y., Sesia, M., and Cand{\`e}s, E.~J.
\newblock Classification with valid and adaptive coverage.
\newblock \emph{arXiv preprint arXiv:2006.02544}, 2020.

\bibitem[Sadinle et~al.(2019)Sadinle, Lei, and Wasserman]{sadinle2019least}
Sadinle, M., Lei, J., and Wasserman, L.
\newblock Least ambiguous set-valued classifiers with bounded error levels.
\newblock \emph{Journal of the American Statistical Association}, 114\penalty0
  (525):\penalty0 223--234, 2019.

\bibitem[Sargent(2001)]{sargent2001comparison}
Sargent, D.~J.
\newblock Comparison of artificial neural networks with other statistical
  approaches: results from medical data sets.
\newblock \emph{Cancer: Interdisciplinary International Journal of the American
  Cancer Society}, 91\penalty0 (S8):\penalty0 1636--1642, 2001.

\bibitem[Shafer \& Vovk(2008)Shafer and Vovk]{shafer2008tutorial}
Shafer, G. and Vovk, V.
\newblock A tutorial on conformal prediction.
\newblock \emph{Journal of Machine Learning Research}, 9\penalty0
  (Mar):\penalty0 371--421, 2008.

\bibitem[Tibshirani \& Foygel(2019)Tibshirani and
  Foygel]{tibshirani2019conformal}
Tibshirani, R. and Foygel, R.
\newblock Conformal prediction under covariate shift.
\newblock \emph{Advances in neural information processing systems}, 2019.

\bibitem[Vovk et~al.(2005)Vovk, Gammerman, and Shafer]{vovk2005algorithmic}
Vovk, V., Gammerman, A., and Shafer, G.
\newblock \emph{Algorithmic learning in a random world}.
\newblock Springer Science \& Business Media, 2005.

\bibitem[Wang et~al.(2019)Wang, Li, and Reddy]{wang2019machine}
Wang, P., Li, Y., and Reddy, C.~K.
\newblock Machine learning for survival analysis: A survey.
\newblock \emph{ACM Computing Surveys (CSUR)}, 51\penalty0 (6):\penalty0 1--36,
  2019.

\bibitem[Wei(1992)]{wei1992accelerated}
Wei, L.-J.
\newblock The accelerated failure time model: a useful alternative to the cox
  regression model in survival analysis.
\newblock \emph{Statistics in medicine}, 11\penalty0 (14-15):\penalty0
  1871--1879, 1992.

\bibitem[Xiang et~al.(2000)Xiang, Lapuerta, Ryutov, Buckley, and
  Azen]{xiang2000comparison}
Xiang, A., Lapuerta, P., Ryutov, A., Buckley, J., and Azen, S.
\newblock Comparison of the performance of neural network methods and cox
  regression for censored survival data.
\newblock \emph{Computational statistics \& data analysis}, 34\penalty0
  (2):\penalty0 243--257, 2000.

\bibitem[Yu et~al.(2011)Yu, Greiner, Lin, and Baracos]{yu2011learning}
Yu, C.-N., Greiner, R., Lin, H.-C., and Baracos, V.
\newblock Learning patient-specific cancer survival distributions as a sequence
  of dependent regressors.
\newblock \emph{Advances in Neural Information Processing Systems},
  24:\penalty0 1845--1853, 2011.

\bibitem[Zhu \& Kosorok(2012)Zhu and Kosorok]{zhu2012recursively}
Zhu, R. and Kosorok, M.~R.
\newblock Recursively imputed survival trees.
\newblock \emph{Journal of the American Statistical Association}, 107\penalty0
  (497):\penalty0 331--340, 2012.

\end{thebibliography}
\bibliographystyle{icml2021}

\appendix
\onecolumn

\newcommand{\hqhi}{\hat{q}_{\alpha_{hi}}}
\newcommand{\hqlo}{\hat{q}_{\alpha_{lo}}}
\newcommand{\qhi}{{q}_{\alpha_{hi}}}
\newcommand{\qlo}{{q}_{\alpha_{lo}}}
\newcommand{\xp}{\xv^\prime}
\newcommand{\tp}{\tv^\prime}

\begin{center}
    \huge{Supplementary Materials}
\end{center}

We complement the omitted proofs in Section~\ref{append: proof}.
We then provide the additional experimental results in Section~\ref{append: experiment}.
Furthermore, in Section~\ref{append: notes}, we give supplementary notes for some statements in the main text.

\section{Proofs}
\label{append: proof}
\subsection{Proof of Theorem~\ref{Thm:robustness}}
\robustness*

Firstly, consider a new sample $\left(\tilde{X}^\prime, \tilde{T}^\prime\right)$ generated from $\tilde{P}_X \times P_{T|X} $, where we assume that $$\mathrm{d}\tilde{P}_X\left(x\right) = \hat{w}\left(x\right) \mathrm{d} {P}_{X|\Delta=1}\left(x\right).$$
As a comparison, due to the definition of $w\left(x\right)$, we have 
$$\mathrm{d}{P}_X\left(x\right) = {w}\left(x\right) \mathrm{d}  {P}_{X|\Delta=1}\left(x\right).$$
We remark that  $\tilde{P}_X\left(x\right)$, ${P}_X\left(x\right)$ are indeed distribution since we assume $\mathbb{E} w\left(X\right) = \mathbb{E} \hat{w}\left(X\right) = 1$, where the expectation is taken over ${P}_{X|\Delta=1}$.

We would first prove that the probability $\tilde{T}^\prime$ falls in the derived confidence interval $\hat{C}^1_n\left(X^\prime\right)$ is larger than $1-\alpha$, where $\alpha$ is the given significance level.
The intuition is that, since the derived $\hat{C}^1_n\left(X^\prime\right)$ is derived based on the estimated weight $\hat{w}\left(x\right)$, the sample $\tilde{T}^\prime$ is then guaranteed to fall in the confidence interval with probability at least $1-\alpha$.

We derive that 
\begin{equation}
\label{eqn: thm1eq1}
\begin{split}
& \mathbb{P}\left(\tilde T^\prime \in \hat{C}^1_n\left(\tilde{X}^\prime\right) | \mathcal{Z}_{tr}\right) \\
=& \mathbb{P}\left(\tilde T^\prime \leq T^u\left(\tilde{X}^\prime\right) | \mathcal{Z}_{tr}\right) \\
=& \mathbb{P} \left(V\left(\tilde{X}^\prime, \tilde T^\prime\right) \leq V\left(\tilde{X}^\prime, T^u\left(\tilde{X}^\prime\right) | \mathcal{Z}_{tr} \right)\right)\\
{\geq}& \mathbb{P} \left(V\left(\tilde{X}^\prime, \tilde T^\prime\right) \leq \operatorname{Quantile} \left(1-\alpha; \sum_{i=1}^n \hat{p}_i \delta_{v_i}+\hat{p}_\infty \delta_{\infty}\right) | \mathcal{Z}_{tr} \right),\\
\end{split} 
\end{equation}
where the last inequality is due to the fact that $V\left(\tilde{X}^\prime, T^u\left(\tilde{X}^\prime\right) \geq \operatorname{Quantile}\right)\left(1-\alpha; \sum_{i=1}^n \hat{p}_i \delta_{v_i}+\hat{p}_\infty \delta_{\infty}\right)$.
We also use the non-decreasing property of $V\left(X, T\right)$ on T.
Furthermore, by Lemma~\ref{lem:equiv}, we can replace the $\delta_{\infty}$ in the quantile term by $\delta_{V\left(\tilde{X}^\prime, \tilde T^\prime\right)}$, therefore,
\begin{equation}
\label{eqn: thm1eq2}
\begin{split} 
&\mathbb{P} \left(V\left(\tilde{X}^\prime, \tilde T^\prime\right) \leq \operatorname{Quantile} \left(1-\alpha; \sum_{i=1}^n \hat{p}_i \delta_{v_i}+\hat{p}_\infty \delta_{\infty}\right) | \mathcal{Z}_{tr} \right) \\
=&\mathbb{P} \left(V\left(\tilde{X}^\prime, \tilde T^\prime\right) \leq \operatorname{Quantile} \left(1-\alpha; \sum_{i=1}^n \hat{p}_i \delta_{v_i}+\hat{p}_\infty \delta_{V\left(\tilde{X}^\prime, \tilde T^\prime\right)}\right) | \mathcal{Z}_{tr} \right).
\end{split} 
\end{equation}
Besides, we know from Lemma~\ref{lem:vprob} that 
$$V\left(\tilde{X}^\prime, \tilde T^\prime\right) |  \mathcal{Z}_{tr}, \mathcal{E} \left(V\right) \sim \sum_{i=1}^n \hat{p}_i \delta_{v_i}+\hat{p}_\infty \delta_{V\left(\tilde{X}^\prime, \tilde T^\prime\right)},$$
therefore, we derive that

\begin{equation}
\label{eqn: thm1eq3}
\begin{split}
& \mathbb{P} \left(V\left(\tilde{X}^\prime, \tilde T^\prime\right) \leq \operatorname{Quantile} \left(1-\alpha; \sum_{i=1}^n \hat{p}_i \delta_{v_i}+\hat{p}_\infty \delta_{V\left(\tilde{X}^\prime, \tilde T^\prime\right)}\right) | \mathcal{Z}_{tr}\right) \\
=& \mathbb{E}_\mathcal{E} \mathbb{P} \left(V\left(\tilde{X}^\prime, \tilde T^\prime\right) \leq \operatorname{Quantile} \left(1-\alpha; \sum_{i=1}^n \hat{p}_i \delta_{v_i}+\hat{p}_\infty \delta_{V\left(\tilde{X}^\prime, \tilde T^\prime\right)}\right) | \mathcal{Z}_{tr}, \ \mathcal{E} \left( V\right) \right) \\
{\geq}& 1-\alpha.
\end{split} 
\end{equation} 

Combining the Equation~\ref{eqn: thm1eq1}, Equation~\ref{eqn: thm1eq2} and Equation~\ref{eqn: thm1eq3} leads to 
\begin{equation}
\mathbb{P}\left(\tilde T^\prime \in \hat{C}^1_n\left(\tilde{X}^\prime\right) | \mathcal{Z}_{tr}\right) \geq 1-\alpha.
\end{equation}

Furthermore, we need to transform the above results into random sample $\left(X^\prime, T^\prime\right) \sim \mathrm{d}{P}_X \times \mathrm{d}{P}_{T|X}$.
This directly follows Lemma~\ref{lem:dtv} that
\begin{equation*}
    \left|\mathbb{P}\left(T^\prime \in \hat{C}^1_n\left(X^\prime\right) | \mathcal{Z}_{tr}\right) - \mathbb{P}\left(\tilde{T}^\prime \in \hat{C}^1_n\left(X^\prime\right) | \mathcal{Z}_{tr}\right)\right| \leq d_{TV}\left(P_{X}\times P_{T|X}, \tilde{P}_{X}\times P_{T|X}\right) = d_{TV}\left(P_{X}, \tilde{P}_{X}\right).
\end{equation*}

We can express the total-variation distance between $Q_{X}$ and $\tilde{Q}_{X}$ as 
\begin{equation*}
    d_{TV}\left(P_{X}, \tilde{P}_{X}\right) = \frac{1}{2} \int |\hat{w}\left(X\right) \mathrm{d} \mathcal{P}_{X|\Delta=1}\left(X\right) - {w}\left(X\right) \mathrm{d} \mathcal{P}_{X|\Delta=1}\left(x\right)| = \frac{1}{2} \mathbb{E}_{X\sim \mathcal{P}_{X|\Delta=1}}|\hat{w}\left(X\right)- {w}\left(X\right)|.
\end{equation*}

Combine the above results and take expectations on the training set, we conclude that
\begin{equation}
    \begin{split}
        &\mathbb{P}\left( T^\prime \in \hat{C}^1_n\left(X^\prime\right) \right) \\
        =& \mathbb{E}_{\mathcal{Z}_{tr}} \mathbb{P}\left( T^\prime \in \hat{C}^1_n\left(X^\prime\right) | \mathcal{Z}_{tr}\right) \\
        \geq & 1-\alpha-\frac{1}{2} \mathbb{E}|\hat{w}\left(X\right)- {w}\left(X\right)|.
    \end{split}
\end{equation}
where the expectation is taken over the training set space and $X \sim \mathcal{P}_{X|\Delta=1}$

\textbf{Technical Lemmas.}
In this part, we give some technical lemmas used in the proof.
We first introduce Lemma~\ref{lem:equiv} which is commonly used in conformal inference.
By Lemma~\ref{lem:equiv}, we can use $\delta_\infty$ to replace $\delta_{V\left(\tilde{X}^\prime, \tilde T^\prime\right)}$ without changing the probability.
\begin{lemma}[Equation(2) in Lemma~1 from \citet{barber2019conformal}.]
\label{lem:equiv}
For random variables $v_i\in\mathbb{R}, i\in[n+1]$, let $p_i\in\mathbb{R}, i\in[n+1]$ be the corresponding weights summing to 1. Then for any $\beta \in [0,1]$, we have 
\begin{equation} 
    V\left(\tilde{X}^\prime, \tilde T^\prime\right) \leq \operatorname{Quantile}\left(\beta, \sum_{i=1}^{n} p_i \delta_{v_i} + p_{\infty} \delta_{\infty} \right) \Longleftrightarrow V\left(\tilde{X}^\prime, \tilde T^\prime\right) \leq \operatorname{Quantile}\left(\beta, \sum_{i=1}^{n} p_i \delta_{v_i} + p_{\infty} \delta_{V\left(\tilde{X}^\prime, \tilde T^\prime\right)}\right).
\end{equation}
\end{lemma}

We next introduce Lemma~\ref{lem:vprob} which provides the distribution of the non-conformity score.
\begin{lemma}
[Equation(A.5) from \citet{lei2020conformal}.]
\label{lem:vprob}

\begin{equation} 
    \left(V\left(\tilde{X}^\prime, \tilde T^\prime\right) | \mathcal{E} \left( V\right)=\mathcal{E} \left(V^*\right), \mathcal{Z}_{tr} \right)\sim \sum_{i=1}^{n} \hat{p}_i \delta_{v^*_i} + \hat{p}_{\infty} \delta_{V\left(\tilde{X}^\prime, \tilde T^\prime\right)},
\end{equation}
where $\mathcal{E}\left(V^*\right)$ is the unordered set of $V^* = \left(v_1^*, v_2^*, \dots, v_n^*, V\left(\tilde{X}^\prime, \tilde{T}^\prime\right)\right)$.
\end{lemma}

Thirdly, we introduce Lemma~\ref{lem:dtv} which shows a basic property of the total variance distance.
\begin{lemma}
[Equation(10) from \citet{berrett2020conditional}.]
\label{lem:dtv}
Let $d_{TV}\left(Q_{1X}, Q_{2X}\right)$ denote the total-variation distance between $Q_{1X}$ and $Q_{2X}$, then 
 \begin{equation} 
    d_{TV}\left(Q_{1X}\times P_{T|X}, Q_{2X}\times P_{T|X}\right) = d_{TV}\left(Q_{1X}, Q_{2X}\right).
\end{equation}
\end{lemma}

\subsection{Proof of Theorem~\ref{thm: double robust}}
\double*

The proof under (A1) directly follows the proof of Theorem~\ref{Thm:robustness}.
Firstly, for a new testing point $\left(X^\prime,T^\prime\right) \sim \mathcal{P}_{X} \times \mathcal{P}_{T|X}$, we have 
\begin{equation}
\label{eqn: thm51eq1}
\begin{split}
    &\mathbb{P}\left(\tv^\prime \in \hat{C}_n^2\left(\xv^\prime\right) | \xp\right)\\
    =&\mathbb{P}\left( \tp\in [\hqlo\left(\xp\right)-\eta, \hqhi\left(\xp\right)+\eta] | \xp \right) \\
    =&\mathbb{P}\left( \max \{ \tp - \hqhi\left(\xp\right), \hqlo\left(\xp\right)-\tp \} \leq \eta | \xv \right) \\
    \overset{i}{\geq} & \mathbb{P}\left( \max \{ \tp - \qhi\left(\xp\right), \qlo\left(\xp\right)-\tp \} \leq \eta - H\left(\xp\right) | \xp \right)  \\
    \overset{ii}{\geq} & \mathbb{P}\left( \max \{ \tp - \qhi\left(\xp\right), \qlo\left(\xp\right)-\tp \} \leq -\varepsilon - H\left(\xp\right) | \xp \right) - \mathbb{P}\left(\eta<-\varepsilon\right) \\
    \overset{iii}{\geq} & \mathbb{P}\left( \max \{ \tp - \qhi\left(\xp\right), \qlo\left(\xp\right)-\tp \} \leq -\varepsilon -  H\left(\xp\right) \mathbb{I}\left( H\left(\xp\right) \leq \varepsilon\right) | \xp \right) - \mathbb{I}\left( H\left(\xp\right) > \varepsilon\right)  -  \mathbb{P}\left(\eta<-\varepsilon\right).
\end{split}    
\end{equation}
Equation (i) follows from Lemma~\ref{Lem: hx}, and Equation (ii) follows from:
\begin{equation*}
    \begin{split}
&\mathbb{P}\left( \max \{ \tp - \qhi\left(\xp\right), \qlo\left(\xp\right)-\tp \} \leq -\varepsilon - H\left(\xp\right) | \xp \right) - \mathbb{P}\left( \max \{ \tp - \hqhi\left(\xp\right), \hqlo\left(\xp\right)-\tp \} \leq \eta - H\left(\xp\right) | \xp \right) \\
\leq& \mathbb{P}\left( \eta - H\left(\xp\right) < \max \{ \tp - \qhi\left(\xp\right), \qlo\left(\xp\right)-\tp \} \leq -\varepsilon - H\left(\xp\right) | \xp \right) \\
\leq & \mathbb{P}\left( \eta - H\left(\xp\right) <-\varepsilon - H\left(\xp\right) | \xp \right)\\
= & \mathbb{P}\left( \eta <-\varepsilon \right).
    \end{split}
\end{equation*}
Equation~(iii) can be derived simply based on the discussion on the value of $\mathbb{I}\left( H\left(\xp\right) \leq \varepsilon\right)$.

By Assumption (B3), since $-\varepsilon - H\left(\xp\right) \mathbb{I}\left( H\left(\xp\right) \leq \varepsilon\right) \leq -2 \varepsilon$, when $ \varepsilon \leq M_2+\gamma$, we have
\begin{equation}
\label{eqn: thm51eq2}
    \begin{split}
        &\mathbb{P}\left( \max \{ \tp - \qhi\left(\xp\right), \qlo\left(\xp\right)-\tp \} \leq -\varepsilon -  H\left(\xp\right) \mathbb{I}\left( H\left(\xp\right) \leq \varepsilon\right) | \xp \right)\\
        \geq & \mathbb{P}\left( \max \{ \tp - \qhi\left(\xp\right), \qlo\left(\xp\right)-\tp \} \leq 0 | \xp \right) - b_2 \left(\varepsilon +  H\left(\xp\right) \mathbb{I}\left( H\left(\xp\right) \leq \varepsilon\right)\right) \\
        \geq & \mathbb{P}\left( \max \{ \tp - \qhi\left(\xp\right), \qlo\left(\xp\right)-\tp \} \leq 0 | \xp \right) - b_2 \left(\varepsilon +  H\left(\xp\right)\right) \\
        \geq & 1-\alpha - b_2 \left(\varepsilon +  H\left(\xp\right)\right).
    \end{split}
\end{equation}

Combining Eqn~\ref{eqn: thm51eq1} with Eqn~\ref{eqn: thm51eq2} and taking expectations over $\xp$, we have:
\begin{equation}
    \begin{split}
    &\mathbb{P}\left(\tv^\prime \in \hat{C}_n^2\left(\xv^\prime\right)\right)\\
    \geq & 1-\alpha - b_2 \left(\varepsilon +  \E H\left(\xp\right)\right)  - \mathbb{P}\left( H\left(\xp\right) > \varepsilon\right)  -  \mathbb{P}\left(\eta<-\varepsilon\right)\\
   \overset{\left(i\right)}{\geq} & 1-\alpha - b_2 \left(2M_2+\gamma\right)  -  \mathbb{P}\left(\eta<-\left(M_2 + \gamma\right)\right).
    \end{split} 
\end{equation}
The Equation~(i) holds by taking $\varepsilon = M_2 + \gamma$.
Due to Assumption~(B1), we have $\E H\left(\xp\right) \leq M_2$ and $\mathbb{P}\left( H\left(\xp\right) > \varepsilon\right) = 0$.
Note that $\varepsilon = M_2 + \gamma$ does not break the condition of Equation~\ref{eqn: thm51eq2}.

We next show that $\lim_{n\to \infty} \mathbb{P}\left(\eta<-\left(M_2 + \gamma\right) \right)\leq \frac{16 M_2}{\left(M_2 + \gamma\right)^2 b_1}$.
Firstly, by Lemma~\ref{lem: eta}, we have
\begin{equation*}
\begin{split}
    \lim_{n\to \infty} \mathbb{P}\left(\eta<-\left(M_2 + \gamma\right) \right) 
    \leq \frac{16}{ \left(M_2 + \gamma\right)^2 b_1} \E [\hat{w}\left(X\right) H\left(X\right)]
    \leq \frac{16M_2}{ \left(M_2 + \gamma\right)^2 b_1} \E [\hat{w}\left(X\right)]
    = \frac{16 M_2}{ \left(M_2 + \gamma\right)^2 b_1}.
\end{split}
\end{equation*}
The inequality is due to Assumption~(B1) and $\E [\hat{w}\left(X\right)]=1$.

The proof is done.

\textbf{Technical Lemmas.}
In this part, we give some technical lemmas used in the proof.
The following Lemma~\ref{Lem: hx} shows that the difference between non-conformity score under true quantile $q$ and non-conformity score under estimated quantile $\hat{q}$ is upper bounded by $H\left(\xp\right)$.
\begin{lemma}
\label{Lem: hx}
Under notations in Theorem~\ref{thm: double robust}, we have
$\left|\max \{ \tp - \hqhi\left(\xp\right), \hqlo\left(\xp\right)-\tp \} -\max \{ \tp - \qhi\left(\xp\right), \qlo\left(\xp\right)-\tp \}\right|  \leq H\left(\xp\right)$
\end{lemma}
\textbf{Proof.} 
We investigate the results via situations on the operator $\max$.

If $\max \{ \tp - \hqhi\left(\xp\right), \hqlo\left(\xp\right)-\tp \} = \tp - \hqhi\left(\xp\right)$ and $\max \{ \tp - \qhi\left(\xp\right), \qlo\left(\xp\right)-\tp \} = \tp - \qhi\left(\xp\right)$, the conclusion follows the definition of $H\left(\xp\right)$.

If $\max \{ \tp - \hqhi\left(\xp\right), \hqlo\left(\xp\right)-\tp \} = \tp - \hqhi\left(\xp\right)$ and $\max \{ \tp - \qhi\left(\xp\right), \qlo\left(\xp\right)-\tp \} = \qlo\left(\xp\right)-\tp$, we have
\begin{equation*}
    \begin{split}
  &\max \{ \tp - \hqhi\left(\xp\right), \hqlo\left(\xp\right)-\tp \} - \max \{ \tp - \qhi\left(\xp\right), \qlo\left(\xp\right)-\tp \}\\
  =& \left(\tp - \hqhi\left(\xp\right)\right) -  \left(\qlo\left(\xp\right)-\tp\right) \\
  \leq& \left(\tp - \hqhi\left(\xp\right)\right) - \left(\tp - \qhi\left(\xp\right)\right) \\
  =& \qhi\left(\xp\right) - \hqhi\left(\xp\right) \\
  \leq& H\left(\xp\right).
    \end{split}
\end{equation*}
Similarly,
\begin{equation*}
    \begin{split}
  &\max \{ \tp - \hqhi\left(\xp\right), \hqlo\left(\xp\right)-\tp \} - \max \{ \tp - \qhi\left(\xp\right), \qlo\left(\xp\right)-\tp \}\\
  =& \left(\tp - \hqhi\left(\xp\right)\right) -  \left(\qlo\left(\xp\right)-\tp\right) \\
  \geq&  \left( \hqlo\left(\xp\right)-\tp\right)-  \left(\qlo\left(\xp\right)-\tp\right) \\
  =& \hqlo\left(\xp\right) - \qlo\left(\xp\right) \\
  \geq& -H\left(\xp\right).
    \end{split}
\end{equation*}
Therefore, we have
$$
\left|\max \{ \tp - \hqhi\left(\xp\right), \hqlo\left(\xp\right)-\tp \} - \max \{ \tp - \qhi\left(\xp\right), \qlo\left(\xp\right)-\tp \}\right| \leq H\left(\xp\right).
$$

The other two situations are derived similarly.

We next introduce Lemma~\ref{lem: eta} which provides the upper bounds of the term $\lim_{n\to \infty} \mathbb{P}\left(\eta<-\varepsilon \right)$.
\begin{lemma}
\label{lem: eta}
Under the assumptions in Theorem~\ref{thm: double robust}, the following inequality holds.
$$\lim_{n\to \infty} \mathbb{P}\left(\eta<-\varepsilon \right) \leq \frac{16}{ \varepsilon^2 b_1} \E [\hat{w}\left(X\right) H\left(X\right)].$$
\end{lemma}
\textbf{Proof.}
By combining Equation~(A.10), (A.11), (A.12), (A.13) in \citet{lei2020conformal}, and apply the Assumption~(B2) we have
$$\lim_{n\to \infty} \mathbb{P}\left(\eta<-\varepsilon \right) \leq \mathbb{P}\left(\sum_{i=1}^n \hat{w}\left(X_i\right)H\left(X_i\right) \geq \frac{\varepsilon^2 b_1 n}{16}\right).$$
And by Markov inequality, we have
$$
\mathbb{P}\left(\sum_{i=1}^n \hat{w}\left(X_i\right)H\left(X_i\right) \geq \frac{\varepsilon^2 b_1 n}{16}\right) \leq \frac{16}{\varepsilon^2 b_1 n} \sum_{i=1}^n \E [\hat{w}\left(X_i\right) H\left(X_i\right)] = \frac{16}{\varepsilon^2 b_1} \E [\hat{w}\left(X\right) H\left(X\right)].
$$
The proof is done.

\subsection{Proof of Theorem~\ref{thm: upperbound}}
\double*

We now show the proof of Theorem~\ref{thm: upperbound}.
For a new testing point $\left(X^\prime,T^\prime\right) \sim \mathcal{P}_{X} \times \mathcal{P}_{T|X}$, we have 
\begin{equation}
\label{eqn: thm52eq1}
\begin{split}
    &\mathbb{P}\left(\tv^\prime \in \hat{C}_n^2\left(\xv^\prime\right) | \xp\right)\\
    =&\mathbb{P}\left( \tp\in [\hqlo\left(\xp\right)-\eta, \hqhi\left(\xp\right)+\eta] | \xp \right) \\
    =&\mathbb{P}\left( \max \{ \tp - \hqhi\left(\xp\right), \hqlo\left(\xp\right)-\tp \} \leq \eta | \xv \right) \\
    \overset{i}{\leq} & \mathbb{P}\left( \max \{ \tp - \qhi\left(\xp\right), \qlo\left(\xp\right)-\tp \} \leq \eta + H\left(\xp\right) | \xp \right)  \\
    \overset{ii}{\leq} & \mathbb{P}\left( \max \{ \tp - \qhi\left(\xp\right), \qlo\left(\xp\right)-\tp \} \leq \eta_{q, w} + \varepsilon + H\left(\xp\right) | \xp \right) + \mathbb{P}\left(\eta-\eta_{q, w}>\varepsilon\right) \\
    \overset{iii}{\leq} &\mathbb{P}\left( \max \{ \tp - \qhi\left(\xp\right), \qlo\left(\xp\right)-\tp \} \leq \eta_{q, w} + \varepsilon + H\left(\xp\right)\mathbb{I}\left( H\left(\xp\right) \leq \varepsilon\right) | \xp \right) + \mathbb{I}\left( H\left(\xp\right) > \varepsilon\right) + \mathbb{P}\left(\eta-\eta_{q, w}>\varepsilon\right) .
\end{split}    
\end{equation}
where we denote $\eta_{q,w} = \operatorname{Quantile}\left(1-\alpha; \sum_{i=1}^n p_i \delta_{{V}_i^*} + p_\infty \delta_{{V}^*_\infty} \right)$, and obviously, $\eta_{q,w}=0$ by the definition of the non-conformity score, where ${V}_i^*$ is calculated based on $\qlo\left(\cdot\right)$ and $\qhi\left(\cdot\right)$. 

Equation (i) follows from Lemma~\ref{Lem: hx}, and Equation (ii) follows from:
\begin{equation*}
    \begin{split}
&\mathbb{P}\left( \max \{ \tp - \qhi\left(\xp\right), \qlo\left(\xp\right)-\tp \} \leq \eta+ H\left(\xp\right) | \xp \right) - \mathbb{P}\left( \max \{ \tp - \hqhi\left(\xp\right), \hqlo\left(\xp\right)-\tp \} \leq \eta_{q, w} + \epsilon + H\left(\xp\right) | \xp \right)  \\
\leq& \mathbb{P}\left( \eta_{q, w} + \epsilon + H\left(\xp\right) < \max \{ \tp - \qhi\left(\xp\right), \qlo\left(\xp\right)-\tp \} \leq \eta + H\left(\xp\right) | \xp \right) \\
\leq &\mathbb{P}\left( \eta_{q, w} + \epsilon + H\left(\xp\right) < \eta + H\left(\xp\right) | \xp \right)\\
= & \mathbb{P}\left( \eta -  \eta_{q, w} >\varepsilon \right).
    \end{split}
\end{equation*}
Equation~(iii) can be derived simply based on the discussion on the value of $\mathbb{I}\left( H\left(\xp\right) \leq \varepsilon\right)$.

By Assumption (C4), since $\varepsilon +H\left(\xp\right) \mathbb{I}\left( H\left(\xp\right) \leq \varepsilon\right) \leq 2 \varepsilon$, when $ \varepsilon \leq M_2+\gamma$, we have
\begin{equation}
\label{eqn: thm52eq2}
    \begin{split}
        &\lim_{n\to \infty} \mathbb{P}\left( \max \{ \tp - \qhi\left(\xp\right), \qlo\left(\xp\right)-\tp \} \leq \eta_{q, w} + \varepsilon +  H\left(\xp\right) \mathbb{I}\left( H\left(\xp\right) \leq \varepsilon\right) | \xp \right)\\
        \leq & \lim_{n\to \infty} \mathbb{P}\left( \max \{ \tp - \qhi\left(\xp\right), \qlo\left(\xp\right)-\tp \} \leq \eta_{q, w} | \xp \right) + b_2 \left(\varepsilon +  H\left(\xp\right) \mathbb{I}\left( H\left(\xp\right) \leq \varepsilon\right)\right) \\
        \leq & \lim_{n\to \infty} \mathbb{P}\left( \max \{ \tp - \qhi\left(\xp\right), \qlo\left(\xp\right)-\tp \} \leq 0 | \xp \right) + b_2 \left(\varepsilon +  H\left(\xp\right)\right) \\
        \leq & 1-\alpha + b_2 \left(\varepsilon +  H\left(\xp\right)\right).
    \end{split}
\end{equation}

The last inequality is from Lemma~\ref{QCI}.
Combining Eqn~\ref{eqn: thm52eq1} with Eqn~\ref{eqn: thm52eq2} and taking expectations over $\xp$, we have:
\begin{equation*} 
    \begin{split}
    &\lim_{n\to \infty} \mathbb{P}\left(\tv^\prime \in \hat{C}_n^2\left(\xv^\prime\right)\right)\\
    \leq & 1-\alpha + b_2 \left(\varepsilon +  \E H\left(\xp\right)\right)  + \mathbb{P}\left( H\left(\xp\right) > \varepsilon\right)  + \lim_{n\to \infty}  \mathbb{P}\left(\eta-\eta_{q, w}>\varepsilon\right) .
    \end{split} 
\end{equation*}
By taking $\varepsilon = M_2^\prime + M_1^\prime/K$, and plugging in Assumption~(B1), the above equation is equivalent to 
\begin{equation}
    \begin{split}
    \lim_{n\to \infty} \mathbb{P}\left(\tv^\prime \in \hat{C}_n^2\left(\xv^\prime\right)\right)\leq  1-\alpha + b_2 \left(2 M_2^\prime + M_1^\prime/K \right)  + \lim_{n\to \infty} \mathbb{P}\left(\eta-\eta_{q, w}>M_2^\prime+ M_1^\prime/K\right).
    \end{split} 
\end{equation}

The left is to show that 
\begin{equation}
  \label{eqn: p}
  \lim_{n\to \infty}  \mathbb{P}\left(\eta-\eta_{q, w}>M_2^\prime+ M_1^\prime/K\right) = 0.  
\end{equation}
First we  denote $\eta_{q} = \operatorname{Quantile}\left(1-\alpha; \sum_{i=1}^n \hat{p}_i \delta_{{V}_i^*} + \hat{p}_\infty \delta_{{V}^*_\infty} \right)$ where $\hat{p}_i$ is the estimator of $p_i$.
Equation~\ref{eqn: p} holds by applying Lemma~\ref{lem: etaetaq} and Lemma~\ref{lem: etaqetaqw}.

The proof is done.

\textbf{Technical Lemmas.}
In this part, we give some technical lemmas used in the proof.
We first prove the following Lemma~\ref{lem: etaetaq} which gives an upper bound of $|\eta - \eta_{q}|$.
\begin{lemma}
\label{lem: etaetaq}
Under assumptions of Theorem~\ref{thm: upperbound}, $|\eta - \eta_{q}|\leq M_2^\prime$
\end{lemma}

\textbf{Proof.}
Notice that $\eta$ is the quantile of distribution $\sum_{i=1}^n \hat{p}_i \delta_{{V}_i} + \hat{p}_\infty \delta_{{V}_\infty}$ and $\eta_q$ is the quantile of distribution  $\sum_{i=1}^n \hat{p}_i \delta_{{V}_i^*} + \hat{p}_\infty \delta_{{V}^*_\infty}$, where $V_i$ is calculated based on $\hqhi, \hqlo$.

When $\eta = \eta_q$, the conclusion directly follows.
When $\eta \not = \eta_q$, notice that the two distribution share the same weight $\hat{w}$, there must exist a $V^\prime$ and $V^{\prime\prime}$ such that
\begin{equation*}
    \begin{split}
        {V^\prime}^* \leq \eta, &\ V^\prime \geq \eta_q\\
        {V^{\prime\prime}}^* \geq \eta, &\ V^{\prime\prime} \leq \eta_q.
    \end{split}
\end{equation*}
This leads to the following two inequalities by Assumption~(C2) and Lemma~\ref{lem: eta}.
\begin{equation}
\begin{split}
    \eta - \eta_q &\geq {V^\prime}^* - V^\prime \geq -H\left(X\right) \geq -M_2^\prime\\
    \eta - \eta_q &\leq {V^{\prime\prime}}^* - V^{\prime\prime}  \leq H\left(X\right) \leq M_2^\prime
\end{split}
\end{equation}
Therefore,
$$
|\eta - \eta_q|\leq M_2^\prime.
$$
The proof is done.

We next prove Lemma~\ref{lem: etaqetaqw} which provides an upper bound of $|\eta_{q} - \eta_{q, w}|$.
\begin{lemma}
\label{lem: etaqetaqw}
Under assumptions of Theorem~\ref{thm: upperbound}, $|\eta_{q} - \eta_{q, w}|\leq M_1^\prime/K$
\end{lemma}

\textbf{Proof.} 
WLOG, assume $\eta_{q,w} - \eta_q = \Delta > 0$.
Denote $F\left(\cdot\right) = \sum_{i=1}^n p_i \delta_{{V}_i^*} + p_\infty \delta_{{V}^*_\infty}$ and $G\left(\cdot\right) = \sum_{i=1}^n \hat{p}_i \delta_{{V}_i^*} + \hat{p}_\infty \delta_{{V}^*_\infty}$. By the definition of $\eta_{q}, \eta_{q, w}$, we have
$$
F\left(\eta_{q, w}\right) = G\left(\eta_q\right) = 1-\alpha.
$$
Therefore, on the one hand, by Assumption~(C1) and Assumption $\E {w}\left(X\right)=\E \hat{w}\left(X\right) = 1$
\begin{equation}
\label{Eqn: eta1}
    \begin{split}
    &F\left(\eta_{q, w}\right) - F\left(\eta_q\right) \\
    =&   G\left(\eta_q\right)  - F\left(\eta_q\right)\\
    \leq& \sup_t G\left(\eta_q\right)  - F\left(\eta_q\right)\\
    \leq& \sup_S |\sum_{i\in S} w\left(X_i\right)-\hat{w}\left(x_I\right)|\\
    \leq& M_1^\prime.
    \end{split}
\end{equation}
On the other hand, by Assumption~(C3)
\begin{equation}
\label{Eqn: eta2}
\begin{split}
    &F\left(\eta_{q, w}\right) - F\left(\eta_q\right) \\
    =& F\left(\eta_{q}+\Delta\right) - F\left(\eta_q\right) \\
    \geq& K\Delta.
\end{split}
\end{equation}
Combining Equation~\ref{Eqn: eta1} and Equation~\ref{Eqn: eta2} leads to 
$$
\eta_{q,w} - \eta_q \leq  M_1^\prime/K.
$$
The conclusion directly follows.

Combining Lemma~\ref{lem: etaetaq} and Lemma~\ref{lem: etaqetaqw} leads to the upper bound of $|\eta - \eta_{q, w}|$.
\section{Supplemental Experiment Results}
\label{append: experiment}
All codes are available at \url{https://github.com/thutzr/Cox}.
\subsection{Experiment Process}
\textbf{Data Pre-processing.} There are both numerical features and categorical features in our datasets. We normalize numerical features on both training set and test set. 

\textbf{Process.} In each run, the experiment runs as follows:
\begin{itemize}
    \item We first randomly split 80\% data as the training set while the rest is splitted randomly into calibration set and test set. 
    \item Following \cite{chen2020deep}, we randomly sample 100 data points, which are used to construct prediction intervals, from test set. Denote these points as $\mathcal{X}_{\text{centers}}$.
    \item For each point $x_0 \in \mathcal{X}_{\text{centers}}$:
    \begin{itemize}
        \item We sample 100 data points in test set with respect to sampling probability proportional to $K(x,x_0)$, where $K(\cdot)$ is the Gaussian kernel.
        \item For these 100 test points, we use algorithm 1 and 2 to calculate the predicted survival interval with respect to the given confidence level $\alpha $ and check if the true survival time of each point is included in the predicted interval. The fraction of points that are covered in the calculated confidence interval is the empirical coverage. And the difference of upper confidence band and its lower counterpart is interval length. In our experiments, the upper interval band is likely to be infinity sometimes. We truncate those upper bands to the maximum duration of the according dataset.
    \end{itemize}
\end{itemize}

We run the above procedure for different confidence level $\alpha$s. Results show that our algorithms are robust and effective for different $\alpha$s. In our experiments, $\alpha$ is chosen to be $0.6,0.7,0.8,0.9,0.95$,respectively.

\subsection{Model and Hyperparameters}
We use pycox and PyTorch to implement CoxCC,CoxPH and neural network model respectively. Then we combine them together to Cox-MLP models. We implement a neural network model with three hidden layers, where each layer has 32 hidden nodes. Between each two layers, we use ReLU as the activation function. We apply batch normalization and dropout which drop 10\% nodes at one epoch. Adam is chosen to be the optimizer. In the training process, we feed 80\% data as training data and the rest as validation data. Note that here the total data set (training data + validation data) is the training data mentioned in the main article. Each batch contains 128 data points. We train the network for 512 epochs and the trained model is used as the MLP part in our Cox-MLP model. 

\subsection{Supplemental Results}
The empirical coverage and predicted interval length of different models are listed in \autoref{tab:s_res} and \autoref{tab:m_res}. The results are similar to that of RRNLNPH. It should be noticed that unweighted WCCI has very poor performance on censored data in SUPPORT. Both Kernel and unweighted models variate a lot on length of prediction interval.

\begin{table}
    \centering
    \caption{Model Comparison on SUPPORT}
    \begin{tabular}{ccccccccc}
    \hline 
        Method & \multicolumn{2}{c}{Total} & \multicolumn{2}{c}{Censored} & \multicolumn{2}{c}{Uncensored} & \multicolumn{2}{c}{Interval Length}\\
        \cline{2-9}
         & Mean & Std.  & Mean & Std.  & Mean & Std.  & Mean & Std.\\
         \hline\hline 
         Cox Reg. & 0.999 & 0.001 & / & / & / & / & 2026.68 & 1.03\\
         Random Survival Forest \cite{ishwaran2008random} & 0.981 & 0.003 & / & / & / & / & 1890.11 & 10.75\\
         Nnet-Survival \cite{gensheimer2019scalable} & 0.995 & 0.002 & 0.988 & 0.005 & 0.998 & 0.001 & 1928.57 & 15.58\\
         MTLR \cite{yu2011learning} & 0.994 & 0.003 & 0.986 & 0.006 & 0.998 & 0.002 & 1921.65 & 37.69\\
         CoxPH \citep{katzman2018deepsurv}& 0.981 & 0.002 & 0.992 &  0.004 & 0.975 & 0.002 & 2029.00& 0.10\\
        CoxCC \citep{katzman2018deepsurv}& 0.981 & 0.002 & 0.992 &  0.004 & 0.975 & 0.003 & 2553.32 & 9.43\\
        CoxPH+WCCI & 0.982 & 0.007 & 0.838 & 0.032 & 0.984 & 0.006 & 1940.85 & 16.41\\
        CoxPH+T-SCI & 0.993 & 0.004 & 0.923 & 0.026 & 0.993 & 0.003 & 2001.08 & 10.09\\
        CoxCC+WCCI & 0.980 & 0.008 & 0.829 & 0.047 & 0.983 & 0.008 &1941.50 & 16.24\\ 
        CoxCC+T-SCI & 0.992 & 0.005 & 0.916 & 0.034 & 0.992 & 0.004& 2001.69 & 10.67\\
        CoxPH+WCCI(unweigted) & 0.825 & 0.023 & 0.543 & 0.054 & 0.957 & 0.010 & 988.06 & 50.67\\
        CoxPH+T-SCI(unweighted) & 0.944 & 0.018 & 0.831 & 0.054 & 0.996 & 0.005 & 1712.76 & 83.19\\
        CoxCC+WCCI(unweigted) & 0.823 & 0.020 & 0.539 & 0.042 & 0.956 & 0.009 & 994.91 & 62.72\\
        CoxCC+T-SCI(unweighted) & 0.942 & 0.021 & 0.825 & 0.063 & 0.997 & 0.004 & 1705.41 & 75.66\\
        Kernel \citep{chen2020deep} & 0.988 & 0.020 & 0.936 & 0.178 & 0.988 & 0.038 & 2027.70 & 30.27\\
        \hline 
    \end{tabular}
    \label{tab:s_res}
\end{table}

\begin{table}
    \centering
        \caption{Model Comparison on METABRIC}
   \begin{tabular}{ccccccccc}
    \hline 
        Method & \multicolumn{2}{c}{Total} & \multicolumn{2}{c}{Censored} & \multicolumn{2}{c}{Uncensored} & \multicolumn{2}{c}{Interval Length}\\
        \cline{2-9}
         & Mean & Std.  & Mean & Std.  & Mean & Std.  & Mean & Std.\\
         \hline\hline 
         Cox Reg. & 0.996 & 0.003 & / & / & / & / & 319.84 & 2.78\\
         Random Survival Forest \cite{ishwaran2008random} & 0.997 & 0.002 &/&/ &/&/ & 341.73 & 4.82 \\
         Nnet-Survival \cite{gensheimer2019scalable} & 0.978 & 0.008 & 0.554 & 0.017 & 0.980 & 0.003 & 19.85 & 0.34\\
         MTLR \cite{yu2011learning} & 0.990 & 0.005 & 0.990 & 0.008 & 0.990 & 0.005 & 306.08 & 6.44\\
         CoxPH \citep{katzman2018deepsurv}& 0.995 & 0.003 & 0.994 &  0.005 & 0.995 & 0.005 & 340.20 & 6.85\\
        CoxCC \citep{katzman2018deepsurv}& 0.996 & 0.004 & 0.994  &  0.005 & 0.998 & 0.004 & 344.26 & 6.93\\
        CoxPH+WCCI & 0.977 & 0.011 & 0.948 & 0.021 & 0.985 & 0.009 & 334.14 & 5.33 \\
        CoxPH+T-SCI & 0.986 & 0.009 & 0.970 & 0.016 & 0.989 & 0.008& 340.67 & 2.91\\
        CoxCC+WCCI & 0.973 & 0.014 & 0.942 & 0.027 & 0.980 & 0.012 & 334.28 & 5.36 \\ 
        CoxCC+T-SCI & 0.986 & 0.008 & 0.971 & 0.015 & 0.990 & 0.007 & 340.80 & 2.98\\
        CoxPH+WCCI(unweigted) & 0.946 & 0.031 & 0.910 & 0.055 & 0.972 & 0.021 & 254.56 & 8.38\\
        CoxPH+T-SCI(unweighted) & 0.958 & 0.063 & 0.932 & 0.063 & 0.977 & 0.021 & 261.18 & 9.26\\
        CoxCC+WCCI(unweigted) & 0.946 & 0.031 & 0.904 & 0.053 & 0.968 & 0.019 & 254.35 & 8.27\\
        CoxCC+T-SCI(unweighted) & 0.958 & 0.063 & 0.926 & 0.061 & 0.975 & 0.022 & 261.17 & 8.87\\
        Kernel \citep{chen2020deep}& 0.981 & 0.025 & 0.997 & 0.010 & 0.971 & 0.038 & 337.66 & 20.39\\
        \hline 
    \end{tabular}

    \label{tab:m_res}
\end{table}

\begin{figure}
    \centering
    \begin{subfigure}[SUPPORT]{0.49\textwidth}
    \includegraphics[width=7cm]{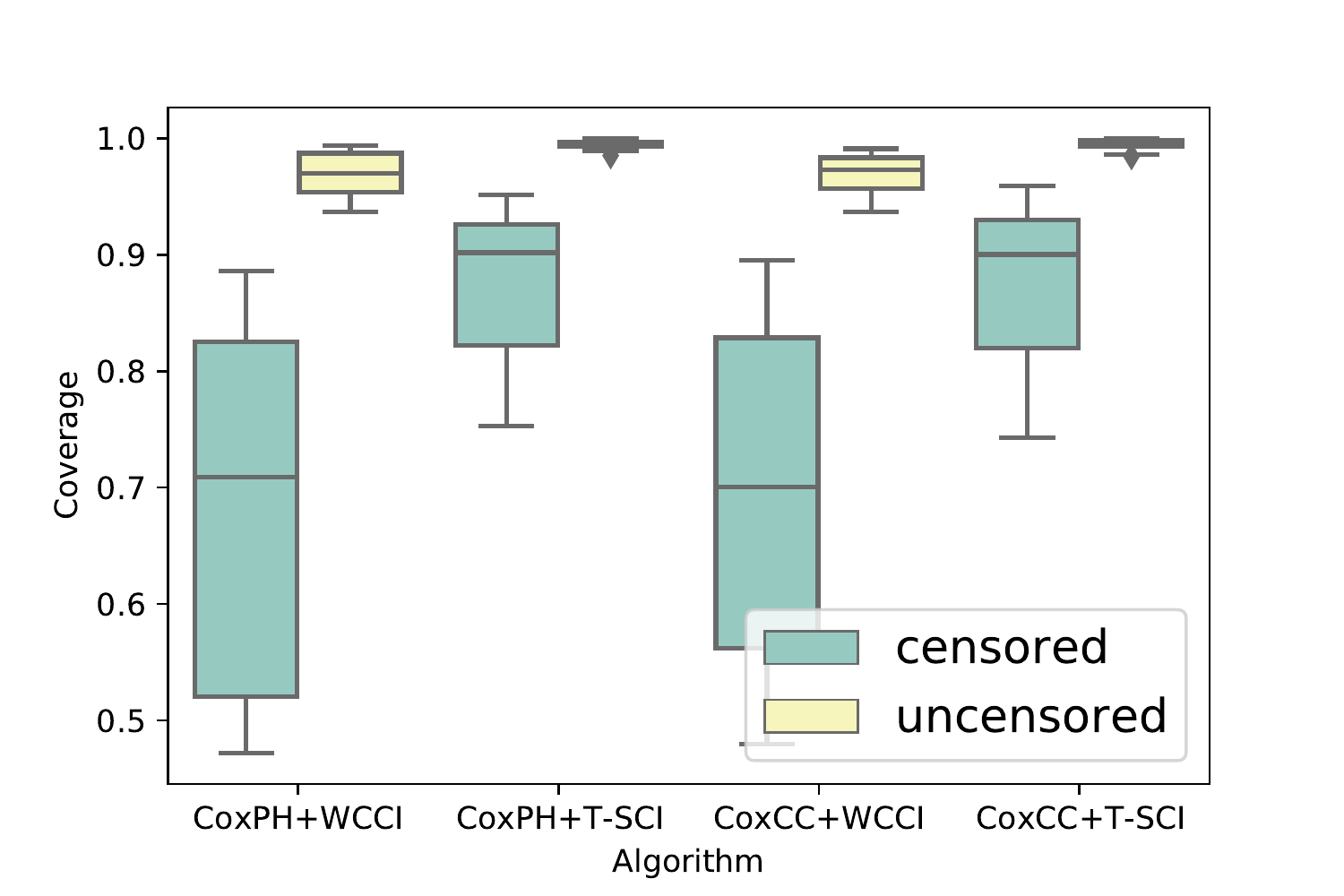}
    \caption{SUPPORT}
    \end{subfigure}
    \begin{subfigure}[METABRIC]{0.49\textwidth}
    \includegraphics[width=7cm]{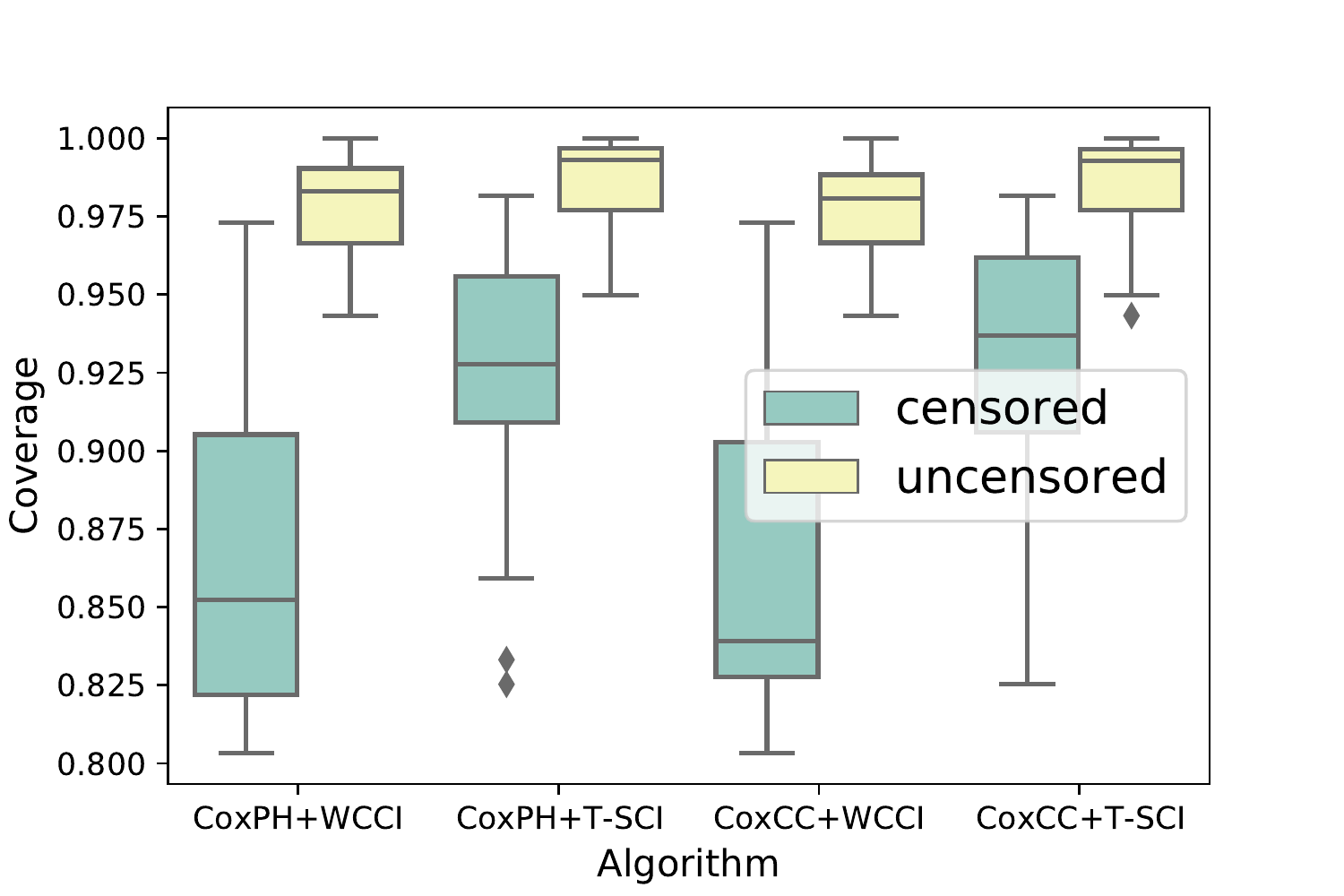}
    \caption{METABRIC}
    \end{subfigure}
    \caption{Empirical Coverage of Censored and Uncensored Data}
    \label{fig:c-uc_c_s}
\end{figure}

Prediction on censored data and uncensored data are compared in \autoref{fig:c-uc_c_s}. Performance on censored data are much better than that of uncensored data. The standard deviation is also larger on uncensored data than censored data. This is consistent with our intuition as we do not have exact information of censored data.

\begin{figure}
    \centering
    \begin{subfigure}[SUPPORT]{0.49\textwidth}
    \includegraphics[width=8cm]{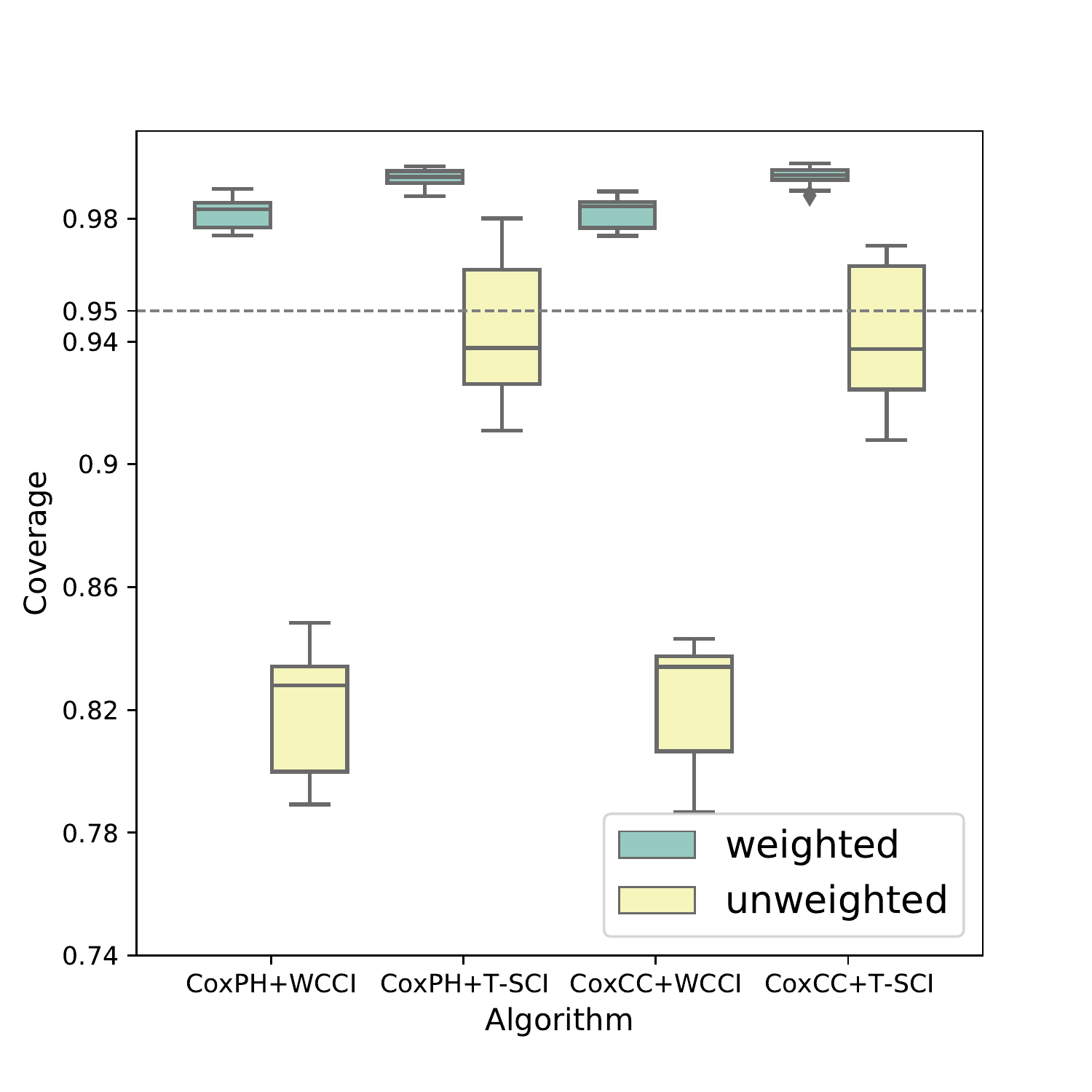}
    \caption{SUPPORT}
    \end{subfigure}
    \begin{subfigure}[METABRIC]{0.49\textwidth}
    \includegraphics[width=8cm]{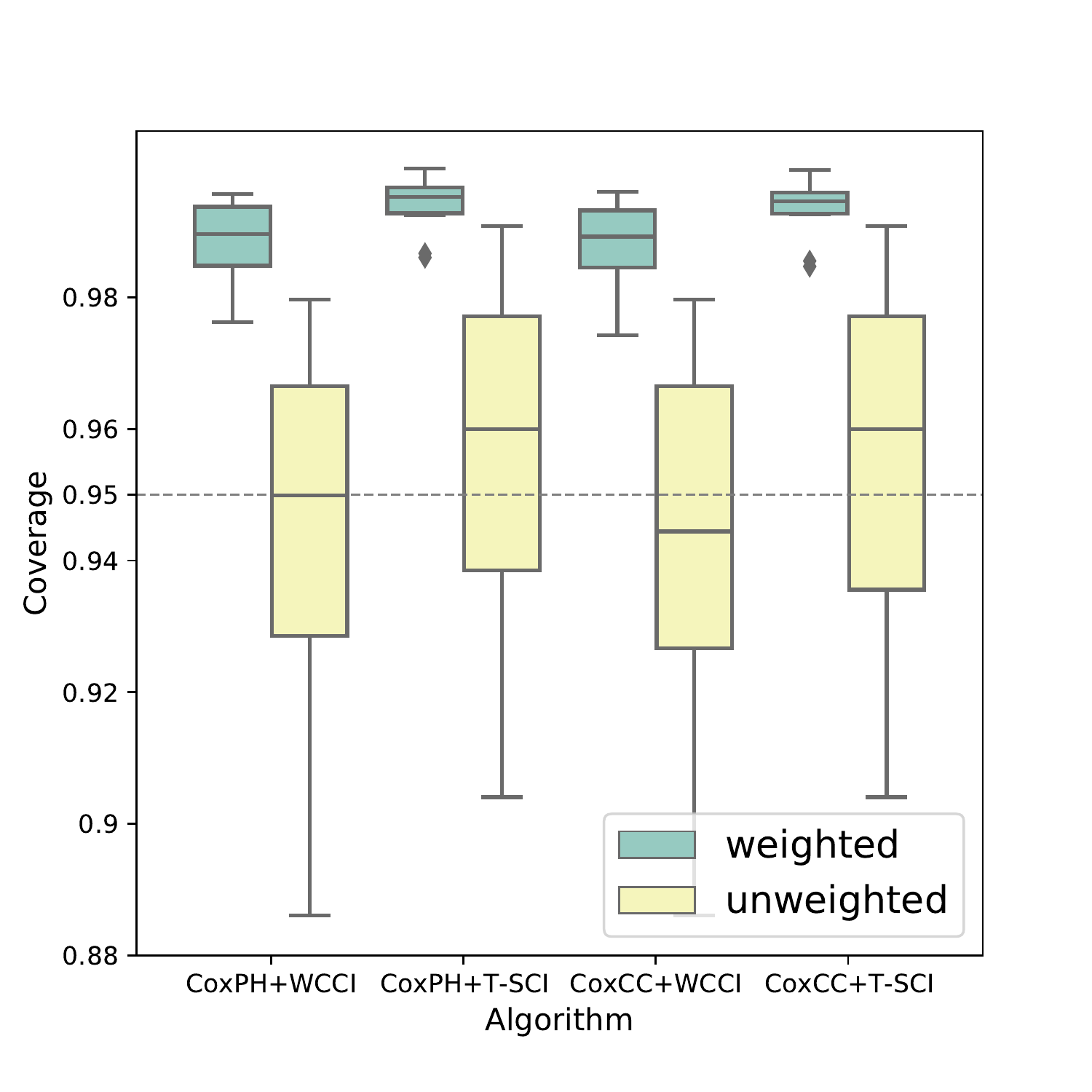}
    \caption{METABRIC}
    \end{subfigure}
    \caption{Empirical Coverage of Weighted and Unweighted Models}
    \label{fig:w-uw_s}
\end{figure}

Comparison on weighted and unweighted models are shown in \autoref{fig:w-uw_s}. Performance of unweighted models are much weaker than weighted models. This shows the effectiveness of weighted conformal inference.

    

\begin{figure}
    \centering
    \begin{subfigure}{0.49\textwidth}
    \includegraphics[height=8cm]{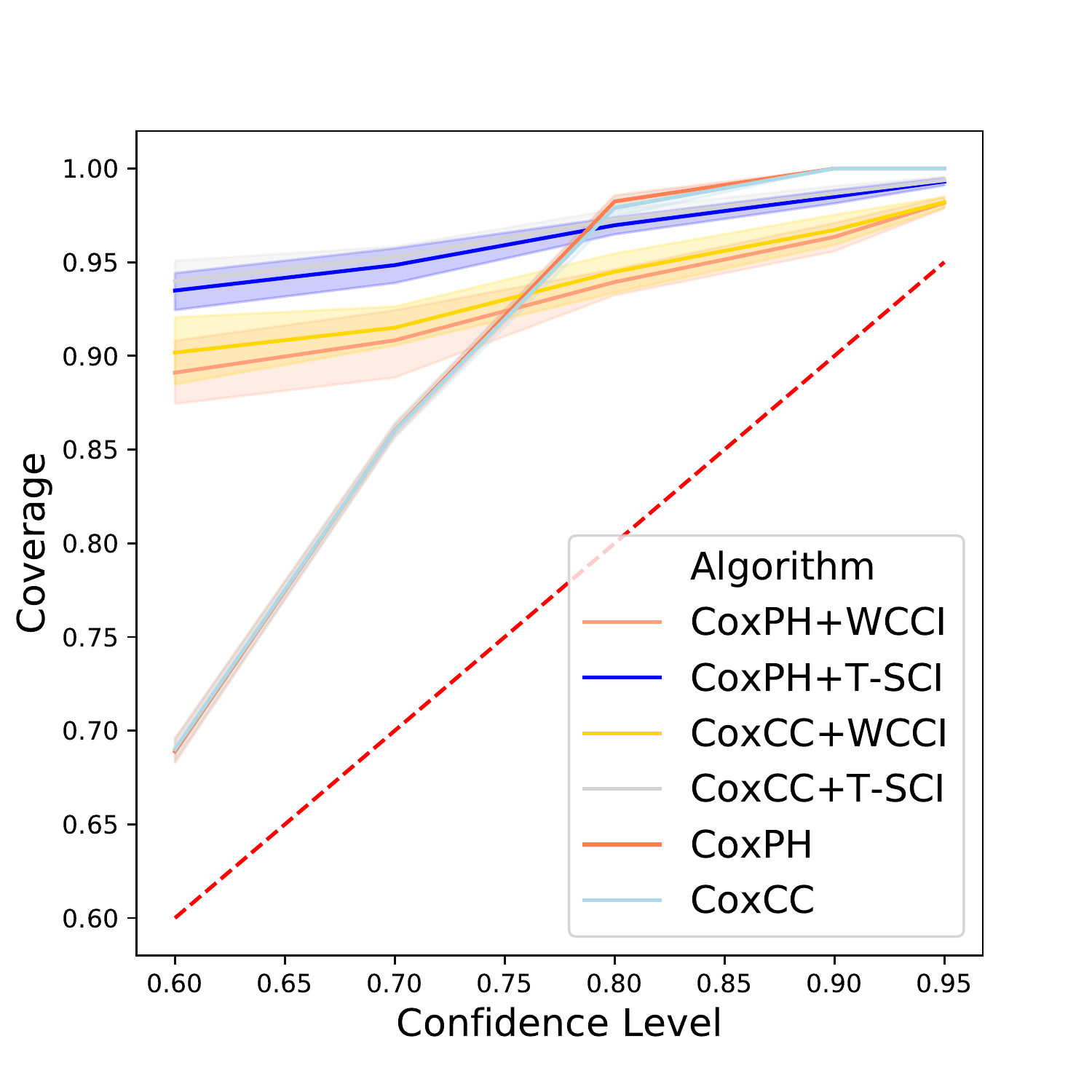}
    \caption{SUPPORT}
    \end{subfigure}
    \begin{subfigure}{0.49\textwidth}
    \includegraphics[height=8cm]{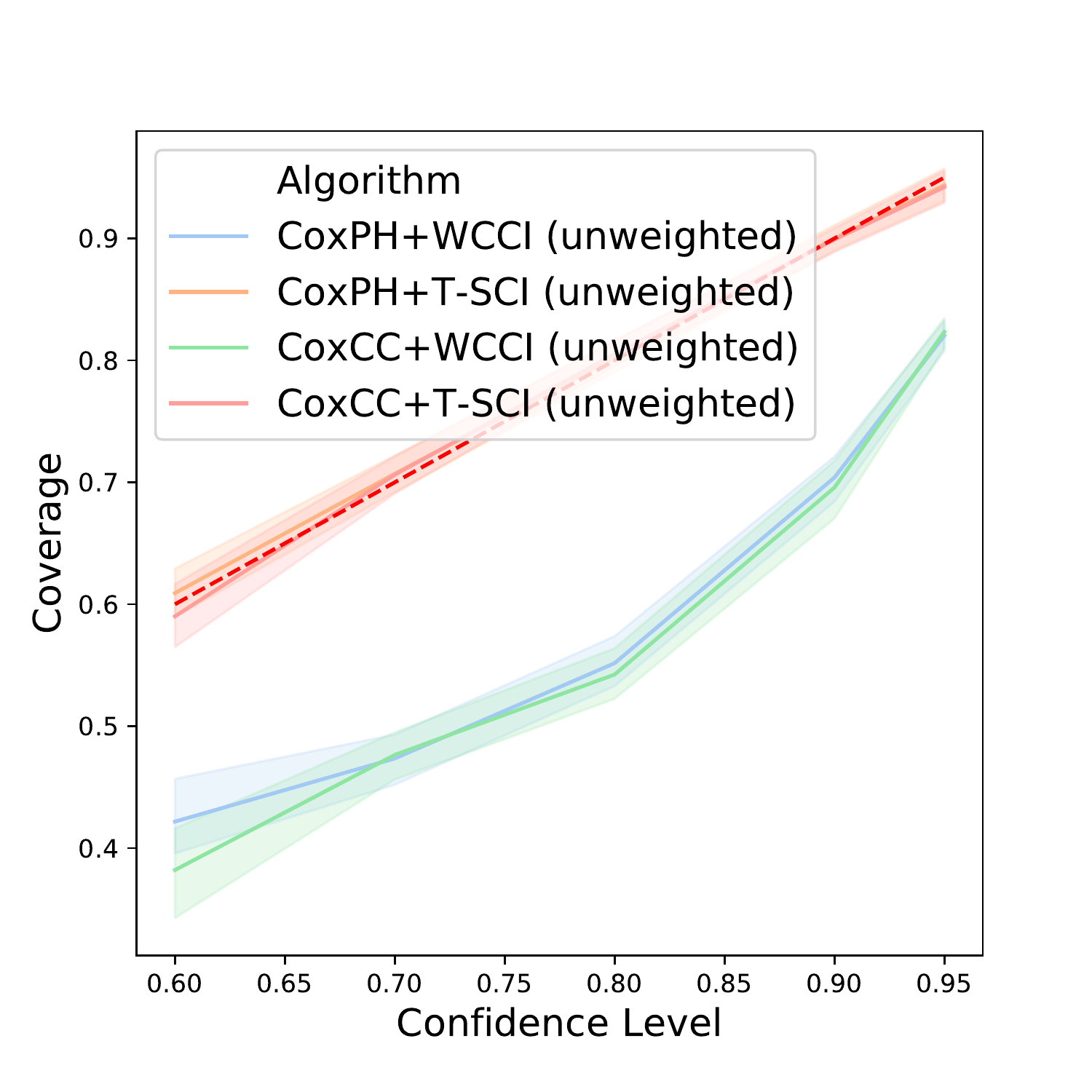}
    \caption{SUPPORT}
    \end{subfigure}
    \begin{subfigure}{0.49\textwidth}
    \includegraphics[height=8cm]{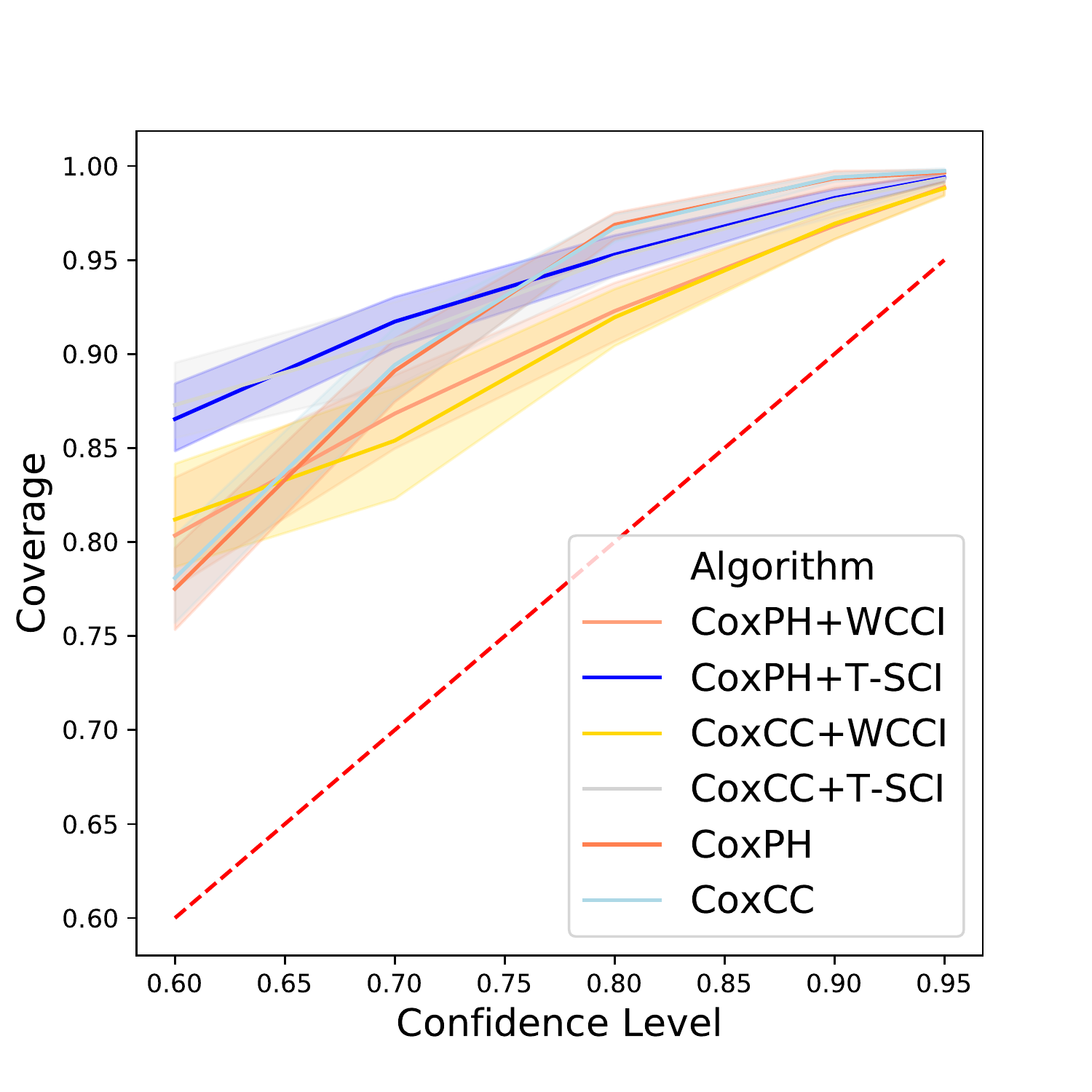}
    \caption{METABRIC}
    \end{subfigure}
    \begin{subfigure}{0.49\textwidth}
    \includegraphics[height=8cm]{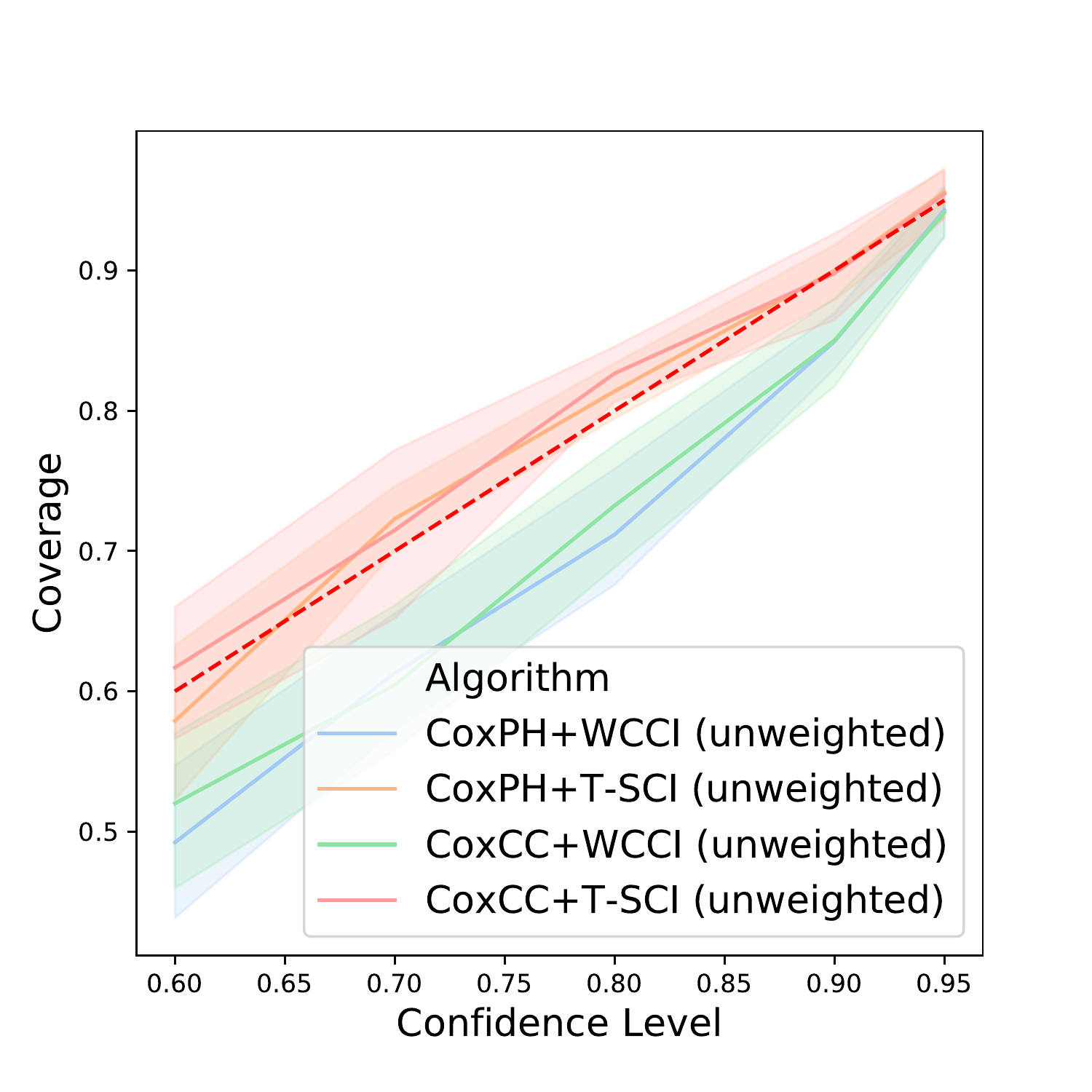}
    \caption{METABRIC}
    \end{subfigure}
    \caption{Different Model's Empirical Coverage of Different $\alpha$(SUPPORT)}
    \label{fig:a_cover_s}
\end{figure}

\begin{figure}[H]
    \centering
    \begin{subfigure}{0.49\textwidth}
    \includegraphics[width=8cm]{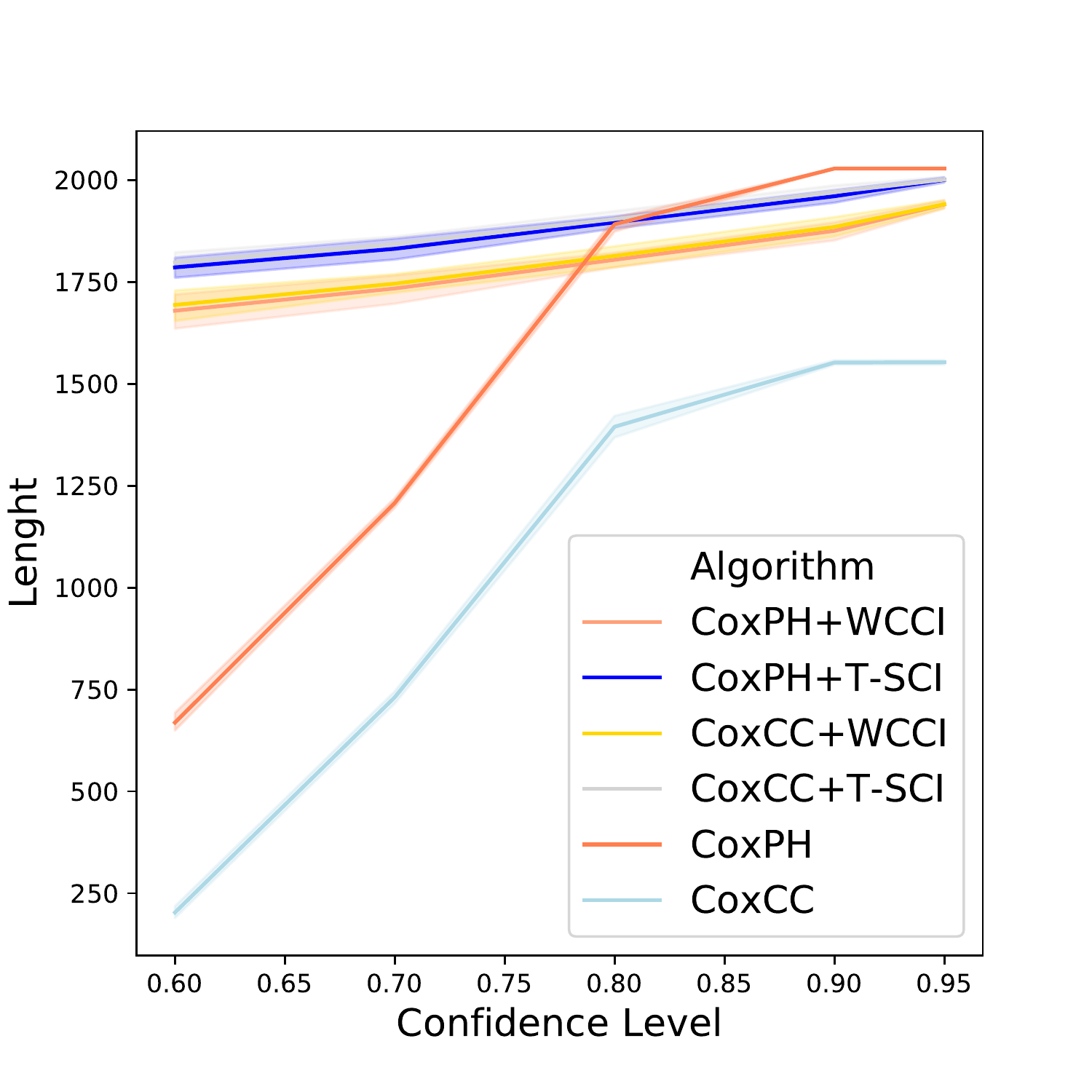}
    \caption{SUPPORT}
    \end{subfigure}
    \begin{subfigure}{0.49\textwidth}
    \includegraphics[width=8cm]{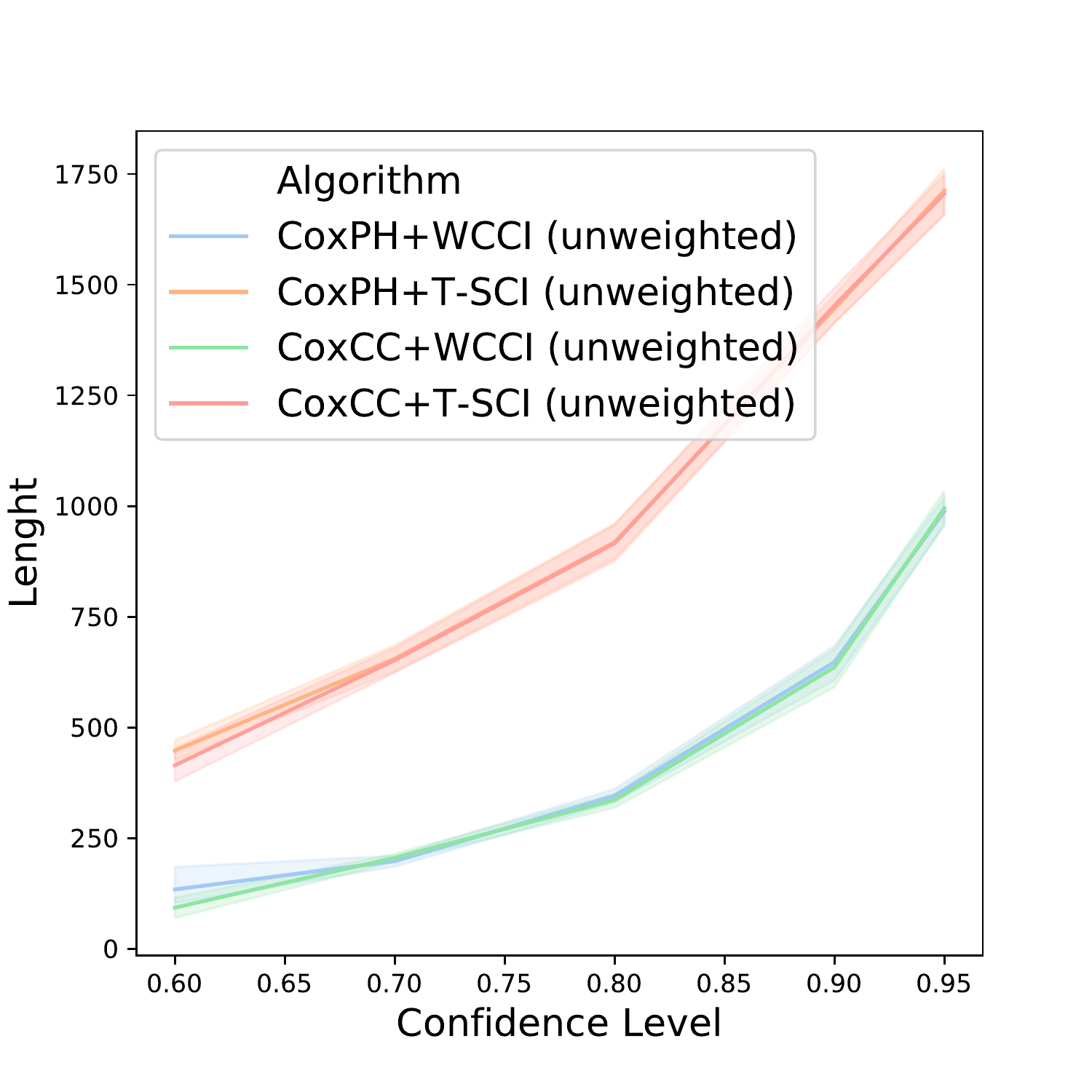}
    \caption{SUPPORT}
    \end{subfigure}
    \begin{subfigure}{0.49\textwidth}
    \includegraphics[width=8cm]{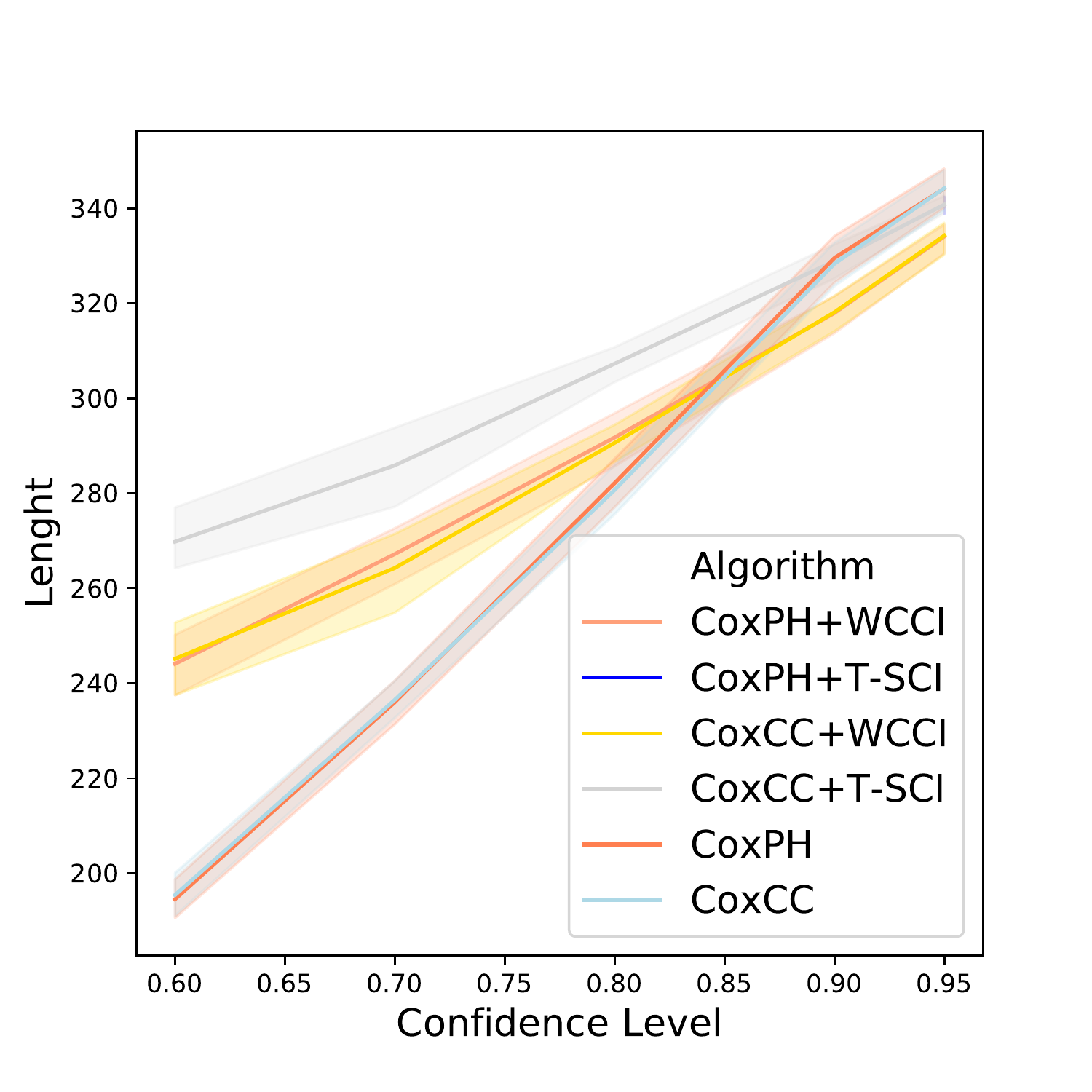}
    \caption{METABRIC}
    \end{subfigure}
    \begin{subfigure}{0.49\textwidth}
    \includegraphics[width=8cm]{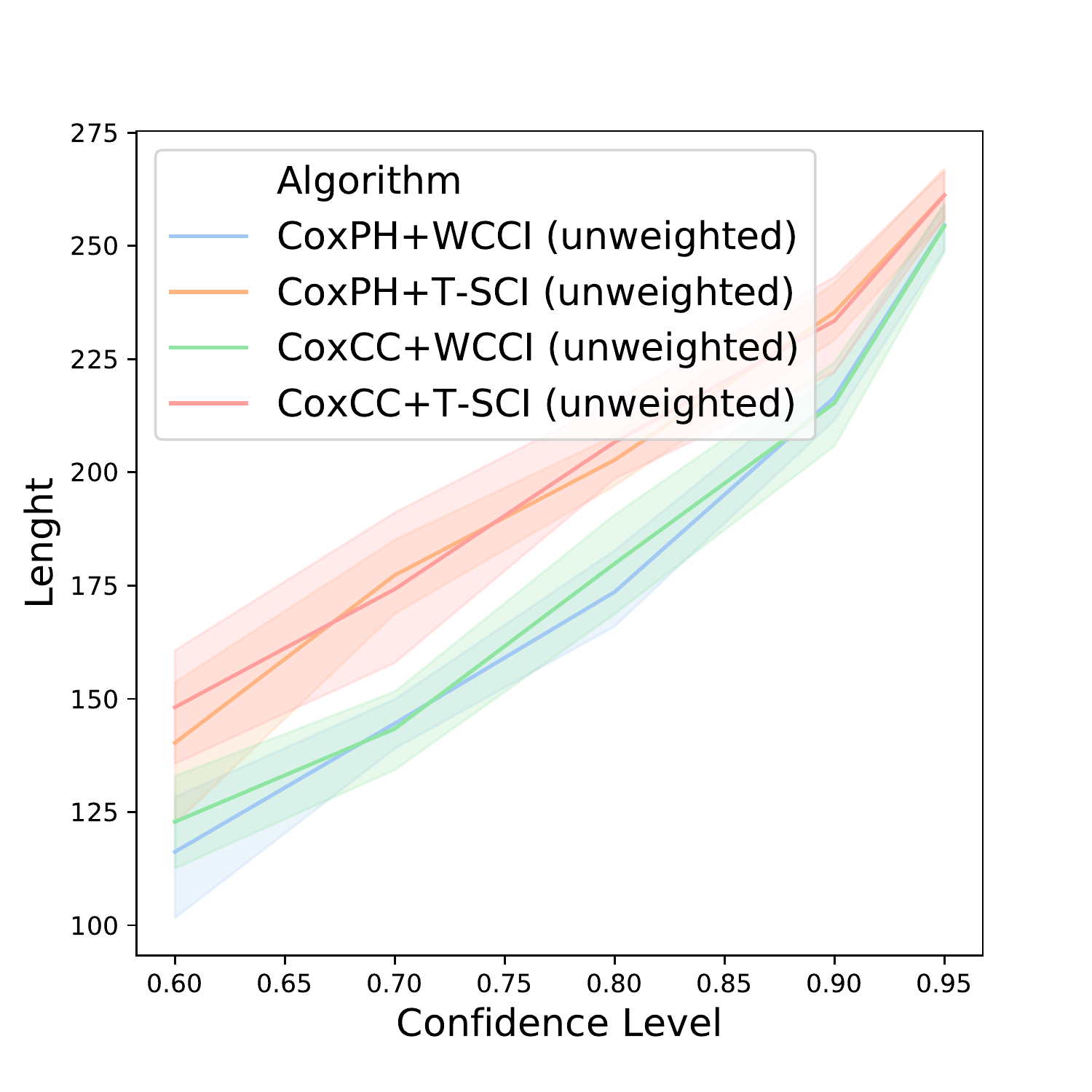}
    \caption{METABRIC}
    \end{subfigure}
    \caption{Different Model's Empirical Coverage of Different $\alpha$}
    \label{fig:a_cover_m}
\end{figure}

We also compare performance on different $\alpha$'s on both SUPPORT  and METABRIC. Results similar with RRNLNPH are shown in \autoref{fig:a_cover_s},
 and \ref{fig:a_cover_m}. The empirical coverage of our algorithms exceed the given confidence level and the standard deviation of prediction is relatively low.

Notice that in Table~\ref{tab:s_res} and Table~\ref{tab:m_res}, censored coverage in CoxPH and CoxCC performs well (around 0.992 in Table~\ref{tab:s_res} and 0.004 in Table~\ref{tab:m_res}).
However, we emphasize that any coverage type could happen due to a lack of theoretical guarantee (extremely large, extremely small, or highly unbalanced, etc.).
The censored coverage of CoxPH happens to be large (0.992) in Table~2 (dataset: SUPPORT) and Table~3 (dataset: METABRIC), just like it happens to be small in Table~1 (0.554, dataset: RRNLNPH).
As a comparison, our newly proposed method (T-SCI) is guaranteed to return a nearly perfect guarantee (all censored coverages are larger than 0.90, and the whole coverages are larger than 0.95).
\section{Supplementary Notes}
\label{append: notes}
We make some supplementary notes in this section.

\subsection{Stability of Non-conformity Score}
We state in Section~4 that the non-conformity score is usually more stable when it is single-peak.
A multi-peak situation implies that samples with different covariate may have different coverage, namely,
$$
\mathbb{P}(T \in C_n(X) | X=x_1) \not = \mathbb{P}(T \in C_n(X) | X=x_2),
$$
where $x_1, x_2$ belong to different peaks.
Therefore, we describe the above formula as ``unstable'' since it provides distinct coverage for distinct groups, although the overall coverage (for the population) is still $1-\alpha$.

The multi-peak phenomenon comes from the fact that we calculate the $1-\alpha$ quantile of $V_i$ based on all samples. 
For example, consider a two-peak distribution where the first group has $1-\alpha$ populations.
Then the algorithm returns zero coverage (probability equal to zero) for the second group and returns one coverage (probability equal to one) for the first group, which causes instability.

\subsection{The Effect of Censoring}
Figure~\ref{fig: biasefficiency} illustrates that ignoring and deleting censoring will indeed cause bias and inefficiency.
We may also consider a more straightforward case to calculate the sample mean of a dataset.
However, the dataset contains censoring issues, where we clip the data to a constant $C$ when the value is larger than $C$.
If we ignore the censoring phenomenon, the new sample mean is smaller than the expected value, leading to \emph{bias}.
If we delete the censoring phenomenon, the new sample mean is also smaller since we delete samples with large values (those deleted samples are always larger than the constant $C$).
Besides, it leads to \emph{inefficiency} since we use fewer samples during the inference.

\subsection{Strong Ignorability Assumption}
Strong ignorability assumption stands for 
$$\tv \ci \iv \ | \ \xv.$$
Note that strong ignorability assumption directly leads to the fact:
$$
\mathcal{P}_{\tv | \xv, \iv} = \mathcal{P}_{\tv | \xv}.
$$
since $ \mathcal{P}_{\tv | \xv, \iv} = \mathcal{P}_{\tv, \iv | \xv} / \mathcal{P}_{\iv | \xv} =  \mathcal{P}_{\tv | \xv}
$.
Therefore, we can apply weighted conformal inference which requires a covariate shift.
A similar idea could be found in \citet{lei2020conformal}.

\subsection{Surrogate Empirical Coverage}
In the experiment part, we use empirical coverage to evaluate the algorithm on a synthetic dataset, defined as the fraction of testing points whose survival time falls in the predicted confidence band.

In the real-world dataset, we calculate \emph{surrogate empirical coverage} (SEC) instead of empirical coverage due to the lack of survival time.
SEC calculates the number when the censoring time is no larger than the band's upper bound for censored data.
We emphasize that SEC is the upper bound of EC.
For example, denote the returned confidence band as $\hat{C}{(X_i)} = [T^l(X_i), T^u(X_i)]$ and denote $\mathcal{I}_{test}$ as the training set index, its empirical coverage is defined as:
$$
EC = \frac{1}{|\mathcal{I}_{test}|} \sum_{i\in\mathcal{I}_{test}}\mathbb{I}(T_i \in \hat{C}{(X_i)})
$$
its surrogate empirical coverage is defined as:
$$
SEC = \frac{1}{|\mathcal{I}_{test}|} \sum_{i\in\mathcal{I}_{test}}\left(\mathbb{I}(Y_i \in \hat{C}{(X_i)}, \Delta_i = 1) + \mathbb{I}(Y_i \leq  T^u(X_i), \Delta_i = 0) \right)
$$

\end{document}